\documentclass[11pt]{article}
\usepackage[a4paper,margin=2.5cm]{geometry}
\usepackage[T1]{fontenc}
\usepackage{lmodern}
\usepackage{amsmath,amssymb,amsthm}
\usepackage{graphicx}
\usepackage{booktabs}
\usepackage[dvipsnames]{xcolor}
\usepackage{authblk}
\usepackage[strings]{underscore}

\definecolor{editred}{rgb}{0.85,0.0,0.0}
\definecolor{editgreen}{rgb}{0.0,0.55,0.0}
\definecolor{editblue}{rgb}{0.0,0.25,0.85}
\newif\ifshowedits
\showeditsfalse
\ifshowedits
  \newcommand{\del}[1]{{\color{editred}#1}}

  \newenvironment{reptext}{\color{editgreen}}{}
  \newenvironment{addtext}{\color{editblue}}{}
\else
  \newcommand{\del}[1]{}

  \usepackage{environ}
  \NewEnviron{deltext}{}

\fi

\definecolor{editioannis}{rgb}{0.90,0.45,0.0}
\definecolor{editpanagiotis}{rgb}{0.55,0.0,0.65}
\ifshowedits

  \newenvironment{ioannistext}{\color{editioannis}}{}
  \newenvironment{panagiotistext}{\color{editpanagiotis}}{}
\else

\fi

\usepackage[numbers,sort&compress]{natbib} 
\usepackage{mathtools}
\usepackage{comment}
\usepackage{framed}
\usepackage[linesnumbered,ruled,commentsnumbered,nofillcomment,noend,nosemicolon,norelsize]{algorithm2e}
\usepackage{fancyhdr}
\usepackage{eqnarray}
\usepackage{enumitem}
\usepackage{siunitx}
\usepackage{wrapfig}
\usepackage{subcaption}
\usepackage{multirow}
\usepackage{caption}
\usepackage{etoolbox}
\makeatletter
\AtBeginEnvironment{table}{\appto\@floatboxreset{\small}}
\AtBeginEnvironment{table*}{\appto\@floatboxreset{\small}}
\makeatother
\usepackage[colorlinks=true,allcolors=black]{hyperref}
\usepackage[capitalize,noabbrev]{cleveref}

\newcommand{\defeq}{\vcentcolon=}

\newcommand{\lstart}{\mathit{start}}
\newcommand{\lend}{\mathit{end}}
\newcommand{\lbegin}{\mathit{begin}}
\newcommand{\Consumers}{\mathit{Con}}
\newcommand{\source}{\mathit{source}}
\newcommand{\Alive}{\mathit{Alive}}

\newcommand{\peak}{\mu}
\newcommand{\candidates}{\mathit{ready}}
\newcommand{\canstart}{\mathit{can\_start}}
\newcommand{\leftparents}{\mathit{left\_par}}
\newcommand{\leftconsumers}{\mathit{left\_con}}
\newcommand{\indata}{\mathit{in\_data}}
\newcommand{\outdata}{\mathit{out\_data}}
\newcommand{\initmem}{\mathit{mem^+}}
\newcommand{\minusmem}{\mathit{mem^-}}

\newcommand{\timeitalic}{\mathit{time}}
\newcommand{\mem}{\mathit{mem}}

\newcommand{\idle}{\mathit{idle}}

\newcommand{\minstart}{\mathit{min\_start}}

\newcommand{\minpe}{\mathit{min\_pe}}
\newcommand{\minidle}{\mathit{min\_idle}}
\newcommand{\loaditalic}{\mathit{load}}
\newcommand{\wmax}{W_\mathit{max}}
\newcommand{\lastconsumer}{\mathit{last\_con}}

\newcommand{\before}{\mathit{before}}

\newcommand{\mastag}{MASTAG}
\newcommand{\lctes}{LCTES23}
\newcommand{\ws}{\textmd{\textsc{WeightedS}}}
\newcommand{\wspp}{\textmd{\textsc{WeightedS++}}}

\theoremstyle{plain}
\newtheorem{theorem}{Theorem}[section]
\newtheorem{proposition}[theorem]{Proposition}

\newtheorem{claim}[theorem]{Claim}
\theoremstyle{definition}
\newtheorem{definition}[theorem]{Definition}

\theoremstyle{remark}

\begin{document}

\title{Memory-Aware Scheduling of Large Task Graphs}

\author[1]{Ioannis~Lamprou}
\author[1,2]{Panagiotis~Xanthopoulos}
\author[1]{Etienne~Filhol}
\author[1]{Zhen~Zhang}
\author[1]{Denis~Barthou}
\author[1]{Yujie~Yuan}
\author[3]{Teng~Su}
\author[2]{Dimitrios~Soudris}
\affil[1]{Distributed \& Parallel Technology Lab, Huawei Technologies France, Paris, France}
\affil[2]{National Technical University of Athens, School~of~Electrical~and~Computer~Engineering, Athens, Greece}
\affil[3]{Huawei Technologies Co., Ltd., Shenzhen, China}
\date{}

\maketitle

\begin{abstract}
In modern applications like deep learning and scientific workflows, memory scarcity is a persistent issue. Current task graph scheduling solutions are either case-specific or computationally expensive, thus creating a need for a general, fast, and efficient approach. In this paper, we consider task graph scheduling on architectures with limited shared memory. We design an algorithm that uses carefully maintained memory consumption information; any schedule it returns provably fits in memory. Our memory-aware adaptive scheduling algorithm runs in polynomial time and requires neither a precomputed sequential schedule nor any search. We compare it to MASTAG, the state of the art for scientific workflows, on 341 task graphs. On the 339 where MASTAG's sequential solver returns a schedule, our main task ordering comes within $0.01\%$ of its memory while being $5.5$ times faster, on median. On 4 and 8 processors, a small portfolio of variants schedules $63\%$ of the instances under a memory limit equal to the sequential peak of MASTAG, and $95\%$ to $96\%$ under one that exceeds it by less than $0.4\%$ on median. The portfolio is, on median, $7$ to $12$ times faster than MASTAG, and $2.9$ to $4.1$ times faster without MASTAG's optional refinement, with makespans within $0.2\%$ to $8.1\%$ of MASTAG's over all limits we test. In training AI foundation models, we achieve up to $1.10\times$ speedup over the default, manually fine-tuned schedule of the MindSpore framework with at most $3.62\%$ more memory, and even memory savings up to $18\%$. On seven training graphs of large language models with up to 115,528 tasks, our Python implementation computes a sequential schedule in at most 33 seconds. With the time limit of its experiments, MASTAG's sequential solver takes 13 minutes to 2.2 hours on five of them and returns no schedule within a day on the two largest.
\end{abstract}


\section{Introduction}\label{sec:intro}

A \textit{task graph} is a powerful representation in parallel programming modeling, able to abstract a variety of applications. 
Scheduling a task graph onto a hardware architecture raises optimization problems regarding performance and resource utilization, which are often intractable in nature.
Nowadays, as the size of applications grows, predominantly in the domain of deep learning \cite{sevilla2022compute}, resources like memory remain scarce \cite{gholami2024ai}.
Employing memory-unaware scheduling may yield solutions which are good in terms of parallelism, yet require more memory than available.

Memory-aware task graph scheduling approaches in the literature focus on special topologies \cite{herrmann2013model, marchal2013scheduling}, specialized priority forcing for graphs with dedicated allocation and release tasks \cite{arras2015list}, and hardware architectures with individual local memories \cite{herrmann2014memory,yao2022memory,kulagina2024mapping,kulagina2025memory}. \citet{marchal2019limiting} add edges so that every parallel schedule respects a memory limit, and \citet{bathie2021dynamic} compute the maximum memory of any schedule on $p$ processors.
For scientific workflows, the state-of-the-art method of \citet{fradet2025parallel} requires the computation of a memory-optimal solution for sequential scheduling, itself an intractable problem. Overall, scalability is questionable when input graphs become too complex, e.g., dense in topology, or of extremely large size, e.g., in modern deep learning applications. Exact and search-based methods are computationally expensive and cannot be applied to use cases where we should stay within the time limit of an interactive compilation process. For example, in our experiments of Section~\ref{sec:experiments-scientific-workflows}, the sequential solver used by \cite{fradet2025parallel} reports a time-out on 48 out of 339 task graphs, and returns no schedule within a day on 2 more; in Section~\ref{sec:experiments-deep-learning}, it returns none within a day on two deep learning graphs, among which one with 115,528 tasks.

\begin{figure}[ht]
    \centering
    \includegraphics[width=0.9\linewidth]{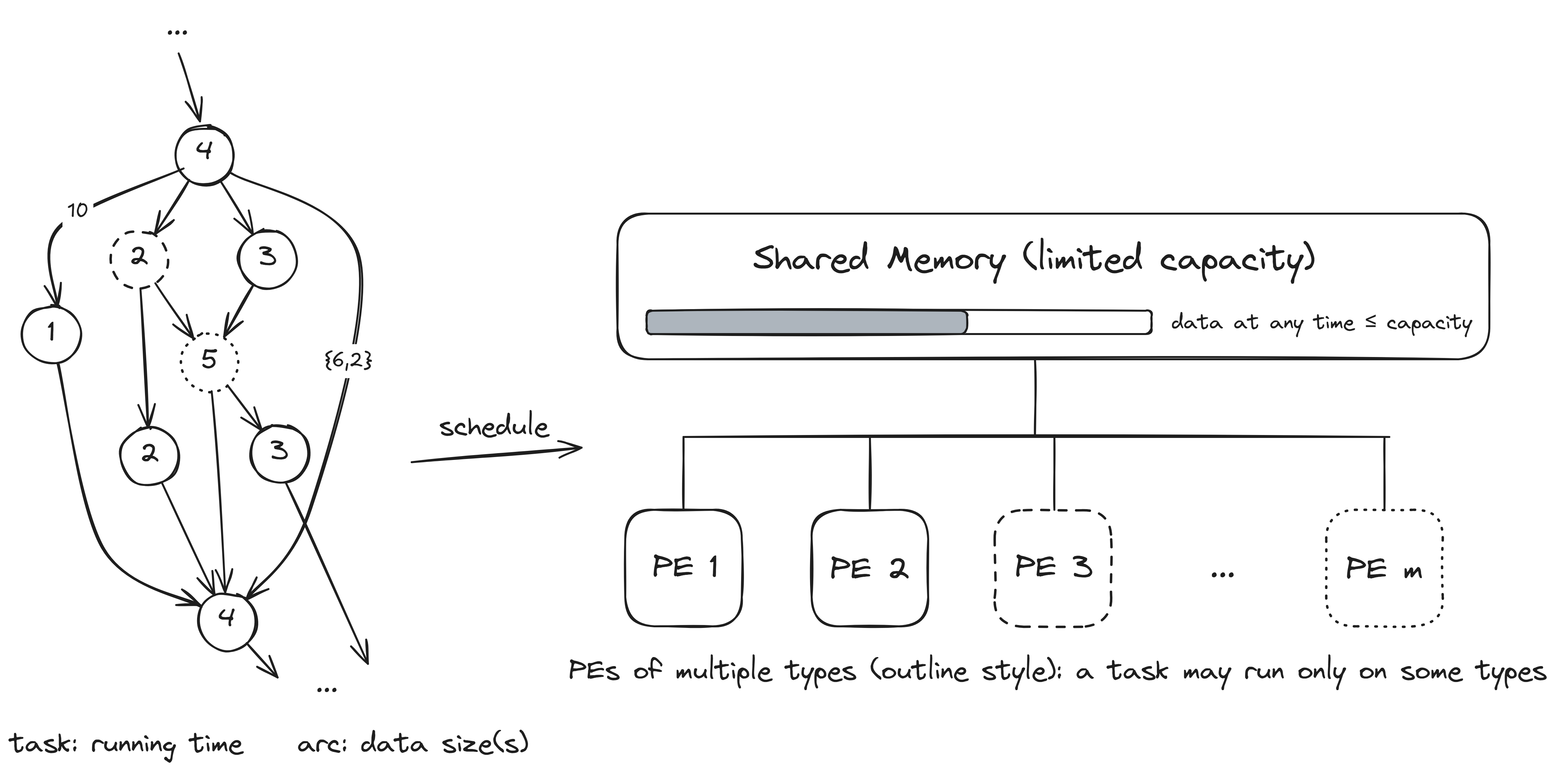}
    \caption{Scheduling a task graph on $m$ processing elements (PEs) of (possibly) different types that share a memory of limited capacity. The number in a task is its running time, and a task may run only on some PE types. An arc carries one or more data items, whose sizes are shown on two arcs. At any time, the data in memory must fit in the capacity.}   \label{fig:hardware}
\end{figure}

Given the above, we identify a need to design a fast and efficient task graph scheduling algorithm to decide safe parallel execution on general architectures comprising multiple \textit{Processing Elements} (PEs) and a shared memory, see Figure~\ref{fig:hardware}. Such an architecture is a first abstraction to model, for example, modern AI accelerators \cite{liao2021ascend}, which include a high bandwidth memory at the top of their memory hierarchy shared by many execution streams.

In a static setting, the whole task graph is known in advance. We aim to find a schedule with a short makespan whose memory usage does not exceed the shared memory limit. To execute it, the schedule is translated into a task launch order with synchronization events between PEs, which keep the execution consistent with the schedule, so that the memory limit also holds at runtime. Since task graphs can be large, e.g., in scientific workflows and deep learning, the scheduling time matters as well. In deep learning frameworks, for example, the schedule is computed during graph compilation, which should stay short.

Similar considerations arise in dynamic settings, where the input is generated on the fly. If little or no information about the task graph is available, a fast algorithmic approach with the aforementioned characteristics can be a powerful tool towards efficient performance and limited out-of-memory errors.

Our algorithm is a greedy list scheduling algorithm \cite{graham1969bounds,kwok1999static,topcuoglu2002performance}: ready tasks are considered one at a time in a priority order, each task is placed on a PE, and no decision is undone later. The difference from classic list scheduling lies in the treatment of memory. When a task is placed, its output data occupy memory from the start of the task until their last consumer finishes, and this finish time is not known yet. Hence, a placement affects memory consumption at all later times, not only during the execution of the task. We keep a memory profile over the whole time horizon and check each placement we try against the memory limit at all times from its start on; the profile never underestimates the memory of the data placed so far, and equals the memory of the schedule at the end. As a result, the memory limit takes part in each decision: if a placement violates the limit, the task is tried in later idle time windows and on other PEs, depending on the rule for choosing the PE, and if it fits nowhere, the next ready task in the priority order is examined instead. A task takes the earliest idle time window that fits, possibly before tasks already placed on the PE (packing), since the profile spans the whole time horizon. Hence, our algorithm is a \textit{memory-aware} and \textit{adaptive} extension of list scheduling: the memory limit has a direct impact on the returned schedule.

The algorithm requires neither a precomputed sequential schedule nor any search. The price to pay is that it may fail to return a schedule, even if one exists within the memory limit. We measure how often this happens. We then run a few variants of the algorithm one after the other, each with a task ordering and a rule for choosing the PE, and keep the schedule with the shortest makespan. This removes most failures once the memory limit leaves a little slack above the peak of a memory-minimizing sequential schedule.

The key contributions of this paper are the following:
\begin{itemize}
    \setlength\itemsep{0.5pt}
    \item We design a memory-aware adaptive list scheduling algorithm for task graphs on PEs of different types with a limited shared memory. The algorithm is parameterized by task and PE sorting methods, and returns either a schedule which provably respects the memory limit, or none. Its running time is polynomial in the size of the input.
    \item We introduce \wspp, a task sorting method which ranks ready tasks by the size of their input data and how long these data have been kept in memory, and breaks ties in favour of tasks which bring other tasks closer to being ready. It is the main ordering of our portfolio, and it greatly reduces the effect of the order of the tasks in the input file on the sequential peak memory.
    \item We compare our algorithm to MASTAG \cite{fradet2025parallel} on 341 task graphs from 17 scientific workflow families, with both tools in \texttt{Python} on the same machine. Like MASTAG, we first minimize the memory of a sequential schedule and then the makespan under a memory limit on several PEs, but with the same algorithm for both stages. For the first, \wspp~comes within $0.01\%$ of the memory of MASTAG's sequential solver while being $5.5$ times faster, on median. For the second, without the need for a precomputed sequential schedule, a portfolio of four runs schedules $63\%$ of the instances under a memory limit equal to the sequential peak of MASTAG, and $95\%$ to $96\%$ under one that exceeds it by less than $0.4\%$ on median. The portfolio is, on median, $7$ to $12$ times faster than MASTAG, and $2.9$ to $4.1$ times faster without MASTAG's optional refinement, with makespans within $0.2\%$ to $8.1\%$ of MASTAG's over all limits we test.
    \item In MindSpore \cite{gptomindspore}, our algorithm gives training speedups up to $1.10\times$ over a manually fine-tuned schedule with at most $3.62\%$ more memory, and memory savings up to $18\%$, on graphs from popular foundation models. On seven training graphs of large language models with up to 115,528 tasks, our \texttt{Python} implementation, with \wspp, computes a sequential schedule in at most 33 seconds. With the time limit of its experiments, MASTAG's sequential solver takes 13 minutes to 2.2 hours on five of them and returns no schedule within a day on the two largest.
\end{itemize}

The \texttt{Python} implementation of our algorithm and the benchmark scripts will be released upon publication to enable reproducibility.

The rest of the paper is organized as follows. In Section~\ref{sec:problems}, we define the scheduling problems. In Section~\ref{sec:algo}, we describe the algorithm, prove it respects the memory limit, bound its running time, and present the task sorting methods. In Section~\ref{sec:experiments}, we present experiments on scientific workflows and in deep learning. We discuss related work in Section~\ref{sec:related} and conclude in Section~\ref{sec:conclusion}.

\section{Problem Definitions}\label{sec:problems}
We define a general problem capturing task graph scheduling on heterogeneous processors with a (physical or logical) limited-size shared memory. We do not model memory reading and writing costs separately; they are included in the task execution cost. In future work, we may expand the definitions to include further cost considerations and hierarchical memory architectures. 

Below, let $\mathbb{N} \defeq \{1, 2, \ldots\}$ denote the set of natural numbers and $\mathbb{N}_0 \defeq \{0\} \cup \mathbb{N}$ denote the set of non-negative integers. For any $x \in \mathbb{N}_0$, we define $[x] \defeq \{0, 1,2, \ldots, x\}$.

The problem input comprises
\begin{itemize}
    \item a \textit{task graph} $G=(N,A)$, that is, a directed acyclic graph where $N$ denotes the set of nodes corresponding to tasks and $A \subseteq \{(n_1, n_2)\,|\, n_1, n_2\in N\}$ the set of directed arcs capturing precedence constraints (data dependencies),
    \item a set of data chunks $D$, where for each data chunk $d \in D$ there is a unique task producing it, namely $\source(d) \in N$, and a set of consumer tasks receiving it, namely $\Consumers(d) \subseteq N$ with $\Consumers(d) \neq \emptyset$, where for each $n \in \Consumers(d)$ there exists an arc $(\source(d), n) \in A$,
    \item an arc-to-data mapping $\alpha: A \rightarrow 2^D \setminus \{\emptyset\}$ indicating the data chunks associated with each arc, where, for $n_1 \neq n_2$, $d \in \alpha((n_1,n_2))$ implies $n_1 = \source(d)$ and $n_2 \in \Consumers(d)$,
    \item a weight function $w: D\rightarrow\mathbb{N}_0$ describing the amount of memory required to store each data chunk,
    \item a set of $m \in \mathbb{N}$ \textit{Processing Elements} (PEs) $P = \{P_0, \ldots, P_{m-1}\}$,
    \item a cost function $c: N \times P \rightarrow \mathbb{N}\cup\{\infty\}$ describing the computational intensity (time) estimated to execute a task on a PE, and
    \item the size of the shared memory $W_{max} \in \mathbb{N}$.
\end{itemize}
In the definitions to follow, we denote the part of the input excluding the shared memory as $I = (G,D,\alpha,w,P,c)$. 

To model processor assignment constraints, when a task $n_i \in N$ is not allowed to execute on $P_j \in P$, we set $c(n_i, P_j) = \infty$. In particular, tasks and PEs may have \textit{types}, e.g., computation and communication, where a task may only execute on PEs of its type.

For any task $n \in N$, there may exist a single self-loop $(n,n)\in A$ associated with the data chunks in $\alpha((n,n))$ in order to model \textit{workspace memory}, that is, memory required  during task execution besides the memory required for input and output data. Workspace data in $\alpha((n,n))$ is associated with arc $(n,n)$ only. Data without consumers, e.g., final outputs, can be modelled as workspace data of their producer.
To simplify the algorithmic descriptions in Section~\ref{sec:algo}, we omit self-loops. Yet, all our algorithms can be easily adjusted to model self-loops, and self-loops are indeed included in our implementation in Section~\ref{sec:experiments}.

Each task $n_i \in N$ \textit{starts} execution on a PE $P_j \in P$ at a discrete time unit $t \in \mathbb{N}_0$, is executed during the interval $[t,t + c(n_i, P_j))$ and \textit{finishes} its execution at time $t + c(n_i, P_j)$. Preemption and task duplication are not allowed.
The problem output is a \textit{schedule}, that is, a decision on when and on which PE to execute each task. A schedule must respect precedence constraints, ensure a PE executes a single task at any time, and execute each task on an allowed PE.

\begin{definition}\label{def:schedule}
    A \textit{schedule} for an input $I=(G,D,\alpha,w,P,c)$ is a pair of functions $\sigma: N \rightarrow \mathbb{N}_0$ and $p: N \rightarrow P$, where $\sigma(n)$ is the start time of task $n$ and $p(n)$ is the PE executing it, such that 
    \begin{itemize}
        \item for $(n_1,n_2)\in A$, with $n_1 \neq n_2$, it holds $\sigma(n_1) + c(n_1, p(n_1)) \le \sigma(n_2)$,
        \item for $n_1, n_2 \in N$, with $n_1 \neq n_2$, if $p(n_1) = p(n_2) = P_j$ holds, then it holds $$[\sigma(n_1), \sigma(n_1) + c(n_1, P_j)) \cap [\sigma(n_2), \sigma(n_2) + c(n_2, P_j)) = \emptyset,$$
        \item and for every $n\in N$ it holds $c(n, p(n)) < \infty$.
    \end{itemize}
    The set of schedules for an input $I$ is denoted by $\Sigma_{I}$.
\end{definition}

\begin{definition}\label{def:makespan}
    The makespan of a schedule $(\sigma, p) \in \Sigma_I$ is its total execution time $$M(\sigma, p) \defeq \max_{n\in N} \{\sigma(n) + c(n, p(n))\}.$$
\end{definition}

Based on the above definitions of schedule and makespan, we first define a standard memory-unaware scheduling problem.

\begin{definition}[GRAPH SCHEDULING (\textbf{GS})]\label{def:graph-schedule}
Given an input $I$, return a schedule $(\sigma^*, p^*) \in \Sigma_{I}$ such that $$M(\sigma^*, p^*) = \min_{(\sigma,p) \in \Sigma_{I}} M(\sigma, p).$$
\end{definition}

GS is NP-hard even in the special case when $P= \{P_0,P_1\}$ and for every $n \in N$, we have $c(n, P_0) = 1$ and $c(n, P_1) = \infty$, or $c(n, P_0) = \infty$ and $c(n,P_1) = 1$ \cite{goyal1976scheduling,bernstein1989complexity}; see also discussion on problem SS9 in \cite{garey1979computers}. For identical PEs, GS is NP-hard already for two PEs and no arcs (problem SS8 in \cite{garey1979computers}).

A solution to GS may be efficient in terms of makespan, that is, in enabling parallelism among PEs. 
Nonetheless, such a solution might require a large amount of memory and lead to out-of-memory problems in case of limited shared memory.
To capture memory needs, we consider \textit{data lifetimes}.
When a task execution begins, storage for all its output data is allocated. A data chunk remains in memory until all its consumer tasks finish execution. This is known as the \textit{produce-before-consume} model \cite{fradet2025graph}; in the \textit{consume-before-produce} model, a task frees its input data before allocating its output data, which may lower the peak memory.

\begin{definition}\label{def:tensor-lifetime}
Given a schedule $(\sigma, p) \in \Sigma_I$, the data lifetime of $d \in D$ is $L_d\defeq [\lbegin_d, \lend_d)$, where $\lbegin_d = \sigma(\source(d))$ and $\lend_d = \max_{n \in \Consumers(d)} \{\sigma(n) + c(n, p(n))\}$. Data
$d \in D$ is \textit{alive} at time $t$ if $t \in L_d$. The set of data alive at time $t$ is denoted as $\Alive(t)$.
\end{definition} 

\begin{definition}\label{def:peak-memory}
Given a schedule $(\sigma, p) \in \Sigma_I$, the peak memory required for parallel execution is
$$W(\sigma,p) \defeq \max_{t \in [M(\sigma,p)]} \sum_{d \in \Alive(t)} w(d).$$
\end{definition}

Note that, in the above definition, it suffices to consider integer times: since all start and finish times are integers, the set of alive data changes only at integer times.

We are now ready to define a \textit{scheduling for memory minimization} problem, regardless of the makespan, and then our main target problem, \textit{scheduling under memory limit}, where we seek to minimize makespan while respecting a given memory limit.

\begin{definition}[GRAPH SCHEDULING FOR MEMORY (\textbf{GS-M})]
\label{def:graph-schedule-memory}
Given an input $I$, return a schedule $(\sigma^*, p^*) \in \Sigma_I$, such that it holds $$W(\sigma^*, p^*) = \min_{(\sigma,p) \in \Sigma_{I}} W(\sigma, p). $$
\end{definition}

\begin{proposition}\label{prop:gsm-hard}
GS-M is NP-hard, even with a single PE, unit costs, and unit data weights.
\end{proposition}
\begin{proof}
We reduce from \textit{Vertex Separation} \cite{lengauer1981black}: given an undirected graph $G=(V,E)$ without isolated vertices and an integer $K$, decide if there is a bijection $f: V \rightarrow \{1, \ldots, |V|\}$ such that $|S_i| \le K$ for all $i$, where $S_i \defeq \{u \in V : f(u) \le i < f(v) \text{ for some } \{u,v\} \in E\}$. We take one task per vertex and one task per edge, with an arc from each of the edge's two endpoints and no other arcs; each vertex task produces one data chunk, consumed by the tasks of its incident edges, and edge tasks produce no data. We show that the minimum peak memory is $\min_f \max_i |S_i| + 1$; hence, there is a schedule with peak memory at most $K+1$ if and only if $G$ has such a bijection.

Fix the order $f$ of the vertex tasks in a schedule, i.e., the task of vertex $v$ is the $f(v)$-th vertex task. For $0 \le i < |V|$, let $v$ be the vertex with $f(v) = i+1$. At the start of the task of $v$, the chunk of each $u \in S_i$ is alive, since $u$ has an edge task that waits for $v$ or a later vertex, and so is the chunk of $v$, which is not in $S_i$; hence the peak is at least $\max_i |S_i| + 1$. Running each edge task right after its later endpoint attains this bound: memory then grows only when a vertex task starts, and only these chunks are alive at that time.
\end{proof}

The reduction above is a short proof for our model. Closely related hardness results are known for register allocation, where a result may also overwrite the register of an operand \cite{sethi1975complete}, and for weighted graphs \cite{jin2023new}.
For series-parallel task graphs, minimizing the peak memory of a sequential schedule can be done in polynomial time \cite{kayaaslan2018scheduling}.

\begin{definition}[GRAPH SCHEDULING WITH  MEMORY LIMIT (\textbf{GS-L})]
\label{def:graph-schedule-memory-opt}
Given an input $I$ and a shared memory size $W_{max}$, return a schedule $(\sigma^*, p^*)\in \Sigma_{I}^{\wmax}$, such that $$M(\sigma^*, p^*) = \min_{(\sigma,p) \in \Sigma_{I}^{\wmax}} M(\sigma, p),$$ where $\Sigma_{I}^{\wmax} = \{(\sigma,p) \in \Sigma_{I}\,|\,W(\sigma,p) \le \wmax\}$, or report that $\Sigma_{I}^{\wmax} = \emptyset$.
\end{definition}

\begin{figure*}[ht]
    \centering
    \includegraphics[width=\linewidth]{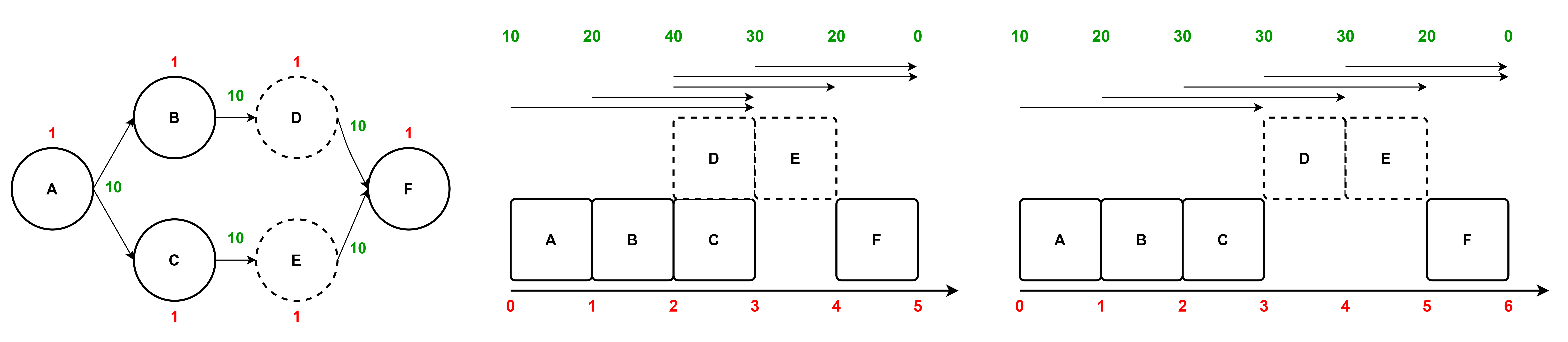}
    \caption{
    On the left, each task can be executed with cost $1$ on a fixed PE (solid for $A, B, C, F$ and dashed for $D,E$) and ships a data chunk of weight $10$ to its children. The schedule in the middle has makespan $5$, but memory peak $40$ while executing $C$ and $D$ in parallel. The schedule on the right has makespan $6$, but memory peak $30$ while executing $C, D, E$ in sequence.}
    \label{fig:example}
\end{figure*}

GS-L is NP-hard, since for $\wmax \ge \sum_{d \in D} w(d)$, it holds $\Sigma_{I}^{\wmax} = \Sigma_{I}$, and GS-L becomes GS. Also, by the reduction above, deciding whether $\Sigma_{I}^{\wmax} \neq \emptyset$ is NP-complete, already on a single PE.
For an intuitive example of the tradeoff between makespan and memory, see Figure~\ref{fig:example}.
We summarize the notation of this section and of Section~\ref{sec:algo} in Table~\ref{tab:notation}.

\begin{table}[ht]
\centering
\small
\caption{Notation}
\label{tab:notation}
\begin{tabular}{ll}
\toprule
Symbol & Meaning \\
\midrule
\multicolumn{2}{l}{\textbf{Problem (Section~\ref{sec:problems})}} \\
$G=(N,A)$ & task graph: tasks $N$, arcs $A$ (data dependencies) \\
$D$, $w(d)$ & data chunks, memory size of chunk $d$ \\
$\alpha((n_1,n_2))$ & data chunks on arc from $n_1$ to $n_2$\\
$\source(d)$, $\Consumers(d)$ & task producing $d$, tasks consuming $d$ \\
$P=\{P_0,\ldots,P_{m-1}\}$ & processing elements (PEs) \\
$c(n,P_j)$ & execution time of task $n$ on $P_j$ ($\infty$ if not allowed) \\
$\wmax$ & memory limit (shared memory size) \\
$(\sigma,p)$ & schedule: start time $\sigma(n)$ and PE $p(n)$ of each task \\
$M(\sigma,p)$, $W(\sigma,p)$ & makespan and peak memory of a schedule \\
$\Alive(t)$ & data in memory at time $t$ \\
\midrule
\multicolumn{2}{l}{\textbf{Algorithm (Section~\ref{sec:algo})}} \\
$\peak$ & memory profile: estimated memory at each time a task starts or finishes \\
$\candidates$ & ready tasks: all parents already scheduled \\
$\canstart(n)$ & earliest start of $n$ allowed by its scheduled parents \\
$\initmem(n)$ & memory allocated when $n$ starts (its output data) \\
$\minusmem(n)$ & memory released when $n$ finishes (data whose only consumer is $n$) \\
$\indata(n)$, $\outdata(n)$ & data received, data produced by task $n$ \\
$\leftparents(n)$ & number of parents of $n$ not yet scheduled \\
$\leftconsumers(d)$, $\lastconsumer(d)$ & consumers of $d$ not yet scheduled, consumer of $d$ finishing last \\
$\lend(d)$ & latest finish time among the scheduled consumers of $d$ \\
$\idle(P_j)$, $\loaditalic(P_j)$ & idle time windows and total assigned cost of $P_j$ \\
$\before(t)$ & latest time in $\peak$ not after $t$ \\
\bottomrule
\end{tabular}
\end{table}

\section{Algorithm Design}\label{sec:algo}

Our algorithm is a \textit{memory-aware} and \textit{adaptive} extension of a list scheduler \cite{topcuoglu2002performance,AlEbrahim2017TaskScheduling}. In each iteration, it takes the ready tasks in the order of a \textit{task sorting method} and places the first one that fits, on a PE chosen by a \textit{PE rule}, in the first idle time window allowed by its parents (first-fit).  Before placing a task, the algorithm checks a \textit{memory profile} $\peak$, which estimates the memory over time and never underestimates it. If the placement may exceed $\wmax$, it tries the next window, PE, or ready task, and if no ready task fits, it stops without a schedule. When a task is placed, its output data are added to $\peak$ from its start time, and once all consumers of a data chunk are placed, the chunk is removed from the finish time of the last of them; Figure~\ref{fig:memory-update} shows one such update on the example of Figure~\ref{fig:example}. At the end of each iteration, the list capturing the order in which tasks are scheduled is readjusted as a result of $\wmax$ violations. Introducing backtracking to recover when no ready task fits would transform the algorithm into a search method and is left for future work.

In the experiments, we use two PE rules: \textit{Load} considers PEs in ascending order of load, i.e., of the total cost of the tasks already assigned to the PE, and places the task on the first PE that admits it (Algorithm~\ref{alg:schedule-task-load}), and \textit{ValidStart} places the task on the PE where it can start earliest (Algorithm~\ref{alg:schedule-task-avail-start}). 
In the captions of the algorithms, LOAD and START denote these rules. A pair of a task sorting method and a PE rule fully defines a run.

First-fit packing provides practical benefits, as we normally do not have to consider all idle time windows where the precedence constraints of a task allow it to be placed. Instead, we place the task as soon as possible.

Section~\ref{sec:algo-description} describes the algorithm, Section~\ref{sec:algo-analysis} proves that a returned schedule respects the memory limit, Section~\ref{sec:task-sorting} presents the task sorting methods, including the novel \ws~and \wspp, and Section~\ref{sec:complexity} bounds the running time.

\subsection{Description}\label{sec:algo-description}

\begin{algorithm}[ht]
    \begin{small}
    \KwIn{$I = (G,D,\alpha,w,P,c)$, $\wmax$, task sorting criterion $\mathit{SORT}$, PE rule $\mathit{PE}$.}
    \KwOut{Schedule $(\sigma, p) \in \Sigma_{I}^{\wmax}$, otherwise None.}
    $\peak \leftarrow \{\}, \sigma \leftarrow \{\}, p \leftarrow \{\}$\;
    $\candidates\leftarrow\mathit{Initialize}(I)$\;
    Let $I_{all} = (I, \wmax, \peak, \sigma, p)$\;
    \While{$\candidates\neq\emptyset$}{
        $n \leftarrow$ $\mathit{ScheduleTask}_{\mathit{[SORT,PE]}}(\candidates, I_{all})$\;
        \lIf{$n == $ {\normalfont None}}{
            \hspace{-0.5mm}%
            \textbf{return} None
        }
        $\mathit{AddMemory}(n, I_{all})$\;
        $\mathit{SubtractMemory}(n,I_{all})$\;
        $\candidates \leftarrow \mathit{UpdateReady}(\candidates, n,  I_{all})$\;
    }
    \textbf{return} $(\sigma, p)$\;
    \caption{Memory-Aware Scheduler}
    \label{alg:mem-aware-graph}
    \end{small}
\end{algorithm}

In Algorithm~\ref{alg:mem-aware-graph}, we provide the main routine of our scheduler. First, all necessary initialization takes place (lines 1--2).
To simplify the algorithm description, we pack all necessary input, output, and auxiliary information into $I_{all}$ (line 3) and later pass it as an argument to all key subroutines. Then, a graph traversal loop (lines 4--9) picks a ready task and schedules it, as long as this is possible in terms of the memory limit (line 5), using the PE rule given as input: Load (Algorithm~\ref{alg:schedule-task-load}) or ValidStart (Algorithm~\ref{alg:schedule-task-avail-start}).
If no ready task can be scheduled without violating the memory limit, the algorithm terminates without returning a schedule (line 6).
Required updates in memory estimates and ready tasks are performed (lines 7--9).
Finally, the schedule is returned (line 10). 

For elements in $N, D, P$, we maintain all necessary auxiliary information, which for presentation purposes we assume is accessed by all subroutines, recall Table~\ref{tab:notation}.
For each $n\in N$, we use the following notation. The value $\canstart(n) \in \mathbb{N}_0$ is a lower bound on $\sigma(n)$ due to previous scheduling decisions, and $\leftparents(n) \in \mathbb{N}_0$ is the number of parents of $n$ not scheduled yet. The sets $\indata(n) \subseteq D$ and $\outdata(n) \subseteq D$ are the data received by $n$ and the data produced by $n$, respectively, and $\initmem(n) \in \mathbb{N}_0$ is the total weight of the data in $\outdata(n)$. Finally, $\minusmem(n)\in\mathbb{N}_0$ is the memory released once $n$ finishes, that is, the total weight of the input data of which $n$ is the only consumer. It is initially zero and is updated during the algorithm.
For each $d \in D$, let $\leftconsumers(d) \subseteq N$ denote the consumers of $d$ not scheduled yet, $\lend(d)$ denote the latest finish time among the consumers of $d$ scheduled so far ($0$ if none), and $\lastconsumer(d)$ denote the consumer whose finish time is $\lend(d)$. Once all consumers of $d$ are scheduled, $\lend(d)$ equals $\lend_d$ of Definition~\ref{def:tensor-lifetime}.
Note that we only care about the maximum finish time of the consumers of each data chunk; it does not matter if more than one consumer finishes at the same time.
For each PE $P_j \in P$, let $\idle(P_j)$ denote the list of idle time intervals for $P_j$ in ascending order of start times, and $\loaditalic(P_j) \in\mathbb{N}_0$ denote the sum of task costs currently assigned to $P_j$.
We let $\peak \subseteq \mathbb{N}_0 \times \mathbb{N}_0$ denote a set of time-to-memory pairs, equivalently a map from time to memory, denoting the memory currently estimated to be needed at times when memory changes, that is, when a task start or finish time occurs. We let $\candidates \subseteq N$ denote the set of currently available tasks to be scheduled, that is, tasks for which all parent tasks are scheduled already.
In $\mathit{Initialize}(\cdot)$, to be defined later in Algorithm~\ref{alg:initialize}, 
we initialize ready tasks as the ones with no parents and assign all necessary initial values.

\begin{algorithm}[ht]
    \begin{small}
    \KwOut{Return a scheduled task $n \in N$, else None.}
        \ForEach{$n\in\candidates$ ordered by $\mathit{SORT}$}{
           \ForEach{{\footnotesize$P_j \in P$ :  $c(n, P_j) < \infty$ ordered by ascending $\loaditalic(\cdot)$}}{
            \ForEach{{\footnotesize$[s,e) \in \idle(P_j)$ in ascending order {\normalfont \textbf{where}} $\max(\canstart(n),s) + c(n, P_j) \le e$}}{
                $\lstart \leftarrow \max(s, \canstart(n))$\;
                 \If{{\normalfont \textbf{not}} {\small $\mathit{MemoryViolated}(n, P_j, \lstart, I_{all})$}}{
                        $\sigma(n) \leftarrow \lstart$\;
                        $p(n) \leftarrow P_j$\;
                        $\mathit{UpdateIdle}(n, \lstart, [s, e), I_{all})$\;
                        $\loaditalic(P_j) \leftarrow \loaditalic(P_j) + c(n, P_j)$\;
                        \textbf{return} $n$\;
                }
             }
           }
        }
       \textbf{return} None\; 
    \caption{$\mathit{ScheduleTask}_{\mathit{[SORT,LOAD]}}(\candidates, I_{all})$}
    \label{alg:schedule-task-load}
    \end{small}
\end{algorithm}

In Algorithm~\ref{alg:schedule-task-load}, the ready set is ordered by the sorting criterion (line 1), then the set of allowed PEs is iterated in ascending order of load \cite{hu2010minimising} (line 2), and finally the list of idle time windows is iterated in ascending order of time (line 3), until an idle window is discovered (if any) where the task fits in terms of time and does not violate the memory limit (lines 4--10). Checking the $\canstart$ constraint is necessary to determine whether the task shall be placed at the beginning of the idle window or later (line 4). In case of success, we schedule the task, assign appropriate $(\sigma, p)$ values to it and update the load and idle list of the PE (lines 6--10).

\begin{algorithm}[ht]
\begin{small}
    \KwOut{True if $\wmax$ violated for task $n$ on $P_j$ at  $\lstart$, else False.}

    \ForEach{$(t, \mem) \in \peak$ {\normalfont \textbf{where}} $\lstart \le t < \lstart + c(n, P_j)$}{
        \If{$\mem + \initmem(n) > \wmax$}{
                \textbf{return} True\;
        }
    }
     \ForEach{{$(t, \mem) \in \peak$ {\normalfont \textbf{where}} $t \ge \lstart + c(n, P_j)$}}{
        \hspace{-1.725mm}%
        \If{$\mem + \initmem(n) - \minusmem(n) > \wmax$}{
                \textbf{return} True\;
        }
    }
    \If{$\peak[\lstart] \text{ is undefined}$}{
        \uIf{
        {\footnotesize
        $\before(\lstart) \neq \emptyset$  
        {\normalfont\textbf{and}} $\peak[\before(\lstart)] + \initmem(n) > \wmax$
        }
        }{  
           \textbf{return} True\;
           }
    }
    \If{$\peak[\lstart + c(n, P_j)] \text{ is undefined}$}{
     \uIf{{\footnotesize%
       $\before(\lstart+ c(n, P_j)) \neq \emptyset$
       {\normalfont\textbf{and}} $ \peak[\before(\lstart + c(n, P_j))] + \initmem(n) - \minusmem(n) > \wmax$
        }
        }{  
           \textbf{return} True\;
           }
    }
    \textbf{return} False\;
    
    \caption{$\mathit{MemoryViolated}(n, P_j, \lstart, I_{all})$}
    \label{alg:memory-violated}
    \end{small}
\end{algorithm}

Algorithm~\ref{alg:memory-violated} receives the examined task $n \in N$, PE $P_j \in P$, and potential start time, and returns True if and only if we estimate the memory limit to be violated by the placement of $n$ on $P_j$ at that start time. We assume $\wmax \ge \max_{n\in N}{\initmem(n)}$ to avoid any trivial memory limit violation.
We examine all points stored in $\peak$. Memory consumption at times strictly before the potential start time of $n$ is not affected. Memory consumption at times overlapping with $n$'s potential execution will be increased by $\initmem(n)$ (lines 1--3).
Memory consumption after the potential finish of $n$ will be increased by $\initmem(n) - \minusmem(n)$, since at least $\minusmem(n)$ is released at the finish of $n$ (lines 4--6).
Special care is required in case the potential start and finish times are not yet defined (lines 7--12). Let $\before(t) \defeq \max\{t' \;|\; \peak[t'] \text{ is defined, } t' \le t\}$ denote the maximum time in $\peak$ which is not larger than $t$, where we let $\before(t) = \emptyset$ if such a maximum does not exist. The memory at time $\before(t)$ is used as a basis to determine the memory at the potentially newly defined time unit $t \in \{\lstart, \lstart + c(n,P_j)\}$.

\begin{algorithm}[ht]
    \begin{small}
    \ForEach{$(\timeitalic, \mem) \in \peak$ {\normalfont \textbf{where}} $\timeitalic \ge \sigma(n)$}{
         $\peak[\timeitalic] \leftarrow \mem + \initmem(n)$\;
    }
    \If{$\peak[\sigma(n)] \text{ is undefined}$}{
        \uIf{
        $\exists (\timeitalic, \mem) \in \peak$ {\normalfont \textbf{where}} $\timeitalic < \sigma(n)$
        }{  
           \hspace{-1.075mm}%
           $\peak[\sigma(n)] \leftarrow \peak[\before(\sigma(n))] + \initmem(n)$\;
        }\Else{
           $\peak[\sigma(n)] \leftarrow  \initmem(n)$\;
        }
    }
    \If{$\peak[\sigma(n) + c(n, p(n))] \text{ is undefined}$}{
        $\peak[\sigma(n) + c(n, p(n))] \leftarrow \peak[\before(\sigma(n) + c(n, p(n)))]$\;
    }

    \caption{$\mathit{AddMemory}(n, I_{all})$}
    \label{alg:add-mem}
    \end{small}
\end{algorithm}

Algorithm~\ref{alg:add-mem} is executed after a task $n$ is scheduled in order to introduce the memory consumption incurred by $n$.
For each time unit in $\peak$ not earlier than the start time of $n$, we add the initial memory needs of $n$ (lines 1--2).
In case the start or finish time of $n$ were not in $\peak$, we introduce them appropriately (lines 3--9).
The consumption at $\sigma(n)$ is equal to the memory incurred by $n$ plus the consumption at the maximum time before $n$ starts execution, if any (lines 4--7).
The consumption at finish time $t' \defeq \sigma(n) + c(n, p(n))$ is the same as the consumption at time $\before(t')$, where either $\before(t') = \sigma(n)$ or  $\sigma(n) < \before(t') < \sigma(n) + c(n, p(n))$ (line 9). In either case, $\peak[\before(t')]$ is already up to date by an assignment in line 2, 5, or 7.

\begin{algorithm}[ht]
    \begin{small}
    
    \ForEach{$d \in \indata(n)$}{
    \If{$\sigma(n) + c(n, p(n)) > \lend(d)$}
        { 
            $\lend(d) \leftarrow \sigma(n) + c(n, p(n))$\;
            $\lastconsumer(d) \leftarrow n$\;
        }
        $\leftconsumers(d) \leftarrow \leftconsumers(d) \setminus \{n\}$\;

        \If{$|\leftconsumers(d)| == 0$}{
            \ForEach{$(\timeitalic,\mem)\in\peak$ {\normalfont \textbf{where}} 
            $\timeitalic \ge \lend(d)$
            }
            {
               $\peak[\timeitalic] \leftarrow \peak[\timeitalic] - w(d)$\;
            }
        }
    }
    \ForEach{$d \in \outdata(n)$}{
        $\lend(d) \leftarrow 0$\;
        \If{$\leftconsumers(d) == \{n'\}$ {\normalfont \textbf{where}} $n' \in N$}{
            $\minusmem(n') \leftarrow \minusmem(n') + w(d)$\; 
        }
    }
    \caption{$\mathit{SubtractMemory}(n, I_{all})$}
    \label{alg:sub-memory}
    \end{small}
\end{algorithm}

In Algorithm~\ref{alg:sub-memory}, we consider the data associated with task $n$.
For each incoming data chunk (lines 1--8), we update its lifetime end, in case $n$ is its currently last consumer in time (lines 2--4) and update its set of remaining consumers (line 5).
If no other consumer is left for this data, we are now certain about when the data is to be subtracted. We subtract the weight of the data at each time greater than or equal to the finish time of its last consumer (lines 6--8).
We also consider outgoing data chunks of $n$ (lines 9--12). We initialize their lifetime end to 0 (line 10), while checking if there is a single consumer $n'$ (lines 11--12). If so, we add $w(d)$ to $\minusmem(n')$.

\begin{figure*}[ht]
    \centering
    \includegraphics[width=\linewidth]{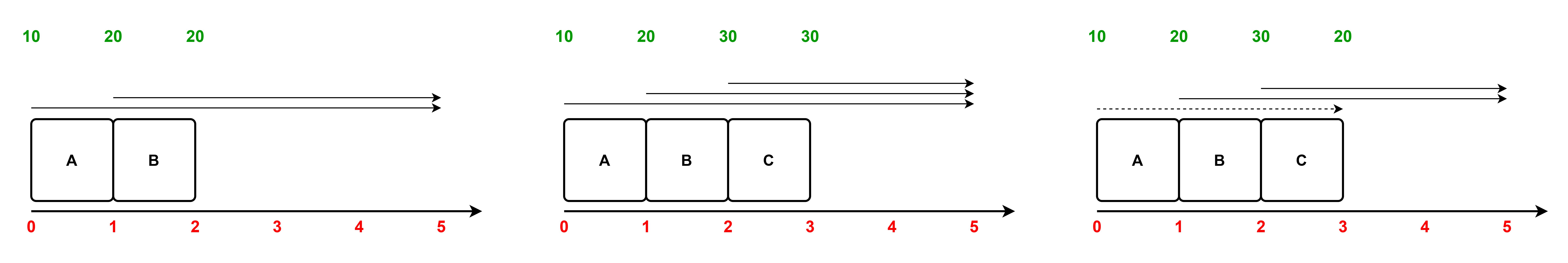}
    \caption{On the left, $A$ and $B$ have already been scheduled. Now, $C$ is selected to be scheduled with $\sigma(C) = 2$. In the middle, $\mathit{AddMemory}(C, I_{all})$ is applied. On the right, $\mathit{SubtractMemory}(C, I_{all})$ is applied, where $C$ is the last consumer of the output of $A$.}
    \label{fig:memory-update}
\end{figure*}

In Figure~\ref{fig:memory-update}, we showcase the effect of Algorithms~\ref{alg:add-mem} and \ref{alg:sub-memory} in the context of the scheduling example in Figure~\ref{fig:example}. 

\paragraph{Remaining subroutines.}

In Algorithm~\ref{alg:initialize}, we perform all necessary initializations of auxiliary information for the scheduler.
Before any task is scheduled, for each $n \in N$ we know it can start at time unit $0$ (line 3), none of its parents has been scheduled (line 4), it is ready to be scheduled if it has no parents (lines 5--6), it is a consumer of all its input data (lines 7--8), where each $\leftconsumers(d)$ is initially empty, it creates an allocation need for all its output data (line 9), and it is unclear whether it is the last consumer of some data (line 10). For a task that is the only consumer of some of its input data, $\minusmem(\cdot)$ is indeed updated in Algorithm~\ref{alg:sub-memory} (lines 11--12). In Algorithm~\ref{alg:initialize}, for each PE, its idle time is originally from $0$ to infinity (line 12) and its load is $0$ (line 13). The set of ready tasks is returned (line 14).

\begin{algorithm}[ht]
    \begin{small}
    \KwOut{Initial set of $\candidates$ tasks.}
    $\candidates \leftarrow \emptyset$\;
    \ForEach{$n \in N$}{
        $\canstart(n) \leftarrow 0$\;
        
        $\leftparents(n) \leftarrow |\{n' \in N\,|\, (n', n) \in A\}|$\;
        \If{$\leftparents(n) == 0$}{
            $\candidates \leftarrow \candidates \cup \{n\}$\;
        }
        \ForEach{$d \in \indata(n)$}{
            $\leftconsumers(d) \leftarrow \leftconsumers(d) \cup \{n\}$\;
        }
        $\initmem(n) \leftarrow \sum_{d \in \outdata(n)} w(d)$\;
        $\minusmem(n) \leftarrow 0$\;
    }
    \ForEach{$P_j \in P$}{
        $\idle({P_j}) \leftarrow [[0, \infty)]$\;
        $\loaditalic(P_j) \leftarrow 0$\;
    }
    \textbf{return} $\candidates$\;
    \caption{$\mathit{Initialize}(I)$}
    \label{alg:initialize}
    \end{small}
\end{algorithm}

\begin{algorithm}[ht]
    \begin{small}
    \KwOut{Return a scheduled task $n \in N$, else None.}
        \ForEach{$n\in\candidates$ ordered by $\mathit{SORT}$}{
           $\minstart \leftarrow \infty$\;
           \ForEach{$P_j \in P$ {\normalfont \textbf{where}} $c(n, P_j) < \infty$}{
            \ForEach{{\footnotesize$[s,e) \in \idle(P_j)$ in ascending order {\normalfont \textbf{where}} $\max(\canstart(n),s) + c(n, P_j) \le e$}}{
                $\lstart \leftarrow \max\{s, \canstart(n)\}$\;
                \If{{\normalfont \textbf{not}} {\small $\mathit{MemoryViolated}(n, P_j, \lstart, I_{all})$}}{
                        \If{$\minstart > \lstart$}{
                         $\minpe \leftarrow P_j$\;
                         $\minstart \leftarrow \lstart$\;
                         $\minidle \leftarrow [s,e)$\;
                        }
                        \textbf{break}\;
                }
             }
           }
           \If{$\minstart < \infty$}{
               $\sigma(n) \leftarrow \minstart$\;
               $p(n) \leftarrow \minpe$\;
               $\mathit{UpdateIdle}(n, \minstart
               ,\minidle, I_{all})$\;
               \textbf{return} $n$\;
           }
        }
       \textbf{return} None\; 
    \caption{$\mathit{ScheduleTask}_{\mathit{[SORT,START]}}(\candidates, I_{all})$}
    \label{alg:schedule-task-avail-start}
    \end{small}
\end{algorithm}

In Algorithm~\ref{alg:schedule-task-avail-start}, we describe an alternative to Algorithm~\ref{alg:schedule-task-load}. Unlike in Algorithm~\ref{alg:schedule-task-load}, where PEs are considered in ascending order of load, here we consider the placement of the task in each allowed PE, e.g., in Algorithm~\ref{alg:schedule-task-avail-start}, line 3, and eventually we decide to schedule the task at a PE such that its start time (Algorithm~\ref{alg:schedule-task-avail-start}, lines 6--11) is minimized. Intuitively, this captures an ``as early as possible'' greedy rationale, in the spirit of \textit{heterogeneous earliest finish time} \cite{topcuoglu2002performance}. It may cost more than Algorithm~\ref{alg:schedule-task-load}, since it examines every allowed PE instead of stopping at the first PE that admits the task. Other more convoluted approaches, where the cost of a task on each PE and the respective memory effect on the partial schedule are taken into account, are left for future work. Testing more potential start points in each idle interval for memory violations would require more complex logic, hence more running time, and is also left for future work.

\begin{algorithm}[ht]
    \begin{small}
    \eIf{$\lstart == s$}{
        Replace $[s, e)$ by $[s + c(n, p(n)), e)$ in $\idle(p(n))$, and let $s \leftarrow s + c(n, p(n))$\;
        \If{$s == e$}{
            Remove $[s, e)$ from $\idle(p(n))$\;
        }
    }{
        Remove $[s, e)$ from $\idle(p(n))$\; 
        Add $[s, \lstart)$ in $\idle(p(n))$\;
        \If{$\lstart + c(n, p(n)) < e$}{
            Add $[\lstart + c(n, p(n)), e)$ in $\idle(p(n))$\;
        }
    }
    
    \caption{$\mathit{UpdateIdle}(n, \lstart, [s, e), I_{all})$}
    \label{alg:update-idle}
    \end{small}
\end{algorithm}

In Algorithm~\ref{alg:update-idle}, 
we modify the idle windows list for the selected PE. If the task is placed at the beginning of the idle window (line 1),
then it suffices to shrink the idle window by shifting its start rightward to the end of the task (line 2).
If the task fills the whole idle window, we remove it from the list (lines 3--4).
In case the task is placed strictly after the start of the idle window (line 5), the idle window is replaced by two new idle windows: one with the idle time before the task starts and one with the idle time after the task finishes, if not empty (lines 6--9).

\begin{algorithm}[ht]
    \begin{small}
    \KwOut{Updated $\candidates$ tasks.}

    $\candidates \leftarrow \candidates \setminus \{n\}$\;
    \ForEach{$n' \in N$ where  $(n, n') \in A$}{
        $\canstart(n') \leftarrow \max\{\canstart(n'),\sigma(n) + c(n, p(n))\}$\;
        $\leftparents(n') \leftarrow \leftparents(n') - 1$\;
        \If{$\leftparents(n') == 0$}{
            $\candidates \leftarrow \candidates \cup \{n'\}$\;
        }
    }
    \textbf{return} $\candidates$\;

    \caption{$\mathit{UpdateReady}(\candidates, n,  I_{all})$}
    \label{alg:update-cand}
    \end{small}
\end{algorithm}

In Algorithm~\ref{alg:update-cand}, the final step of a scheduling iteration,
we update ready tasks by removing the scheduled task (line 1) 
and inserting tasks whose parents have all been scheduled (lines 4--6). 
The $\canstart$ values are updated, so that each child of the scheduled task is only allowed to start after it (line 3), in order to guarantee correctness in terms of precedence constraints.

\subsection{Analysis}\label{sec:algo-analysis}
We prove the algorithm is safe in its memory estimate, in other words, it never underestimates memory consumption.

Intuitively, the profile $\peak$ is a step function, which changes only at start and finish times of scheduled tasks. When a task is scheduled, the memory of its output data is added from its start time until the end of the time horizon, since the consumer finishing last is not known yet. The memory of a data chunk is subtracted, from the finish time of its last consumer onward, only once all its consumers are scheduled. Thus, at any iteration, each data chunk produced so far is counted at least during its whole lifetime, and the profile may only overestimate memory. Once all tasks are scheduled, each data chunk is counted exactly during its lifetime, and the profile equals the memory consumption of the schedule. Since each placement is checked against the profile, the returned schedule respects the memory limit. The formal argument follows.

Let $N_i \subseteq N$, for $i \in \mathbb{N}$, denote the set of the $i$ tasks scheduled in the first $i$ iterations in Algorithm~\ref{alg:mem-aware-graph}, $D_i \defeq \{d \in \outdata(n)\;|\; n \in N_i\}$, and $w(D_i) \defeq \sum_{d \in D_i} w(d)$.
Let $\peak_i[\cdot]$ denote the contents of $\peak[\cdot]$ after the first $i$ iterations and $\before_i(t) \defeq \max\{t' \;|\; \peak_i[t'] \text{ is defined}, t' \le t\}$ denote the max time in $\peak_i$ which is not larger than $t$. 

\begin{claim}\label{claim:no-underestimate-memory} Assume Algorithm~\ref{alg:mem-aware-graph} returns a schedule $(\sigma, p)$.
For each $i \in \mathbb{N}$, where $i \le |N|$, for each time $t \in [M(\sigma, p)]$, it holds $\sum_{d \in \Alive(t) \cap D_i} w(d) \le \peak_i[\before_i(t)]$.
Also, for each time $t \in [M(\sigma, p)]$, it holds $\sum_{d \in \Alive(t)} w(d) = \peak_{|N|}[\before_{|N|}(t) ]$.
\end{claim}
\begin{proof}
For $i \le |N|$ and $t \in [M(\sigma, p)]$, let $$A_i(t) \defeq \{d \in D_i \;|\; \lbegin_d \le t \text{ and } (\lend_d > t \text{ or } \Consumers(d) \not\subseteq N_i)\}$$ denote the data counted in $\peak_i$ at time $t$. Since $d \in \Alive(t)$ means $\lbegin_d \le t < \lend_d$, it holds $\Alive(t) \cap D_i \subseteq A_i(t)$, and it suffices to show $\sum_{d \in A_i(t)} w(d) \le \peak_i[\before_i(t)]$.
The first task $n_1 \in N$ is scheduled at time $0$ on $P_j \in P$, since $\canstart(n_1) = 0$ and all PEs are idle. Hence, from now on, $\peak$ contains the time $0$, and $\before_i(t)$ is defined for all $t$. By Algorithm~\ref{alg:add-mem}, it holds $\peak_1[0] = \peak_1[c(n_1, P_j)] = w(D_1)$. For  $t \in [c(n_1, P_j)-1]$, it holds
$
\sum_{d \in A_1(t)} w(d) = w(D_1) = \peak_1[0] = \peak_1[\before_1(t)]
$, and for $t \in [c(n_1, P_j), M(\sigma, p)]$, it holds
$$
\sum_{d \in A_1(t)} w(d) = \sum_{d \in D_1} w(d) =w(D_1) = \peak_1[c(n_1, P_j)] = \peak_1[\before_1(t)].
$$

Inductively, we assume $\sum_{d \in A_{i-1}(t)} w(d) \le \peak_{i-1}[\before_{i-1}(t)]$. The $i$-th task $n_i \in N$ is now scheduled on $P_j \in P$ for some $j$. By Algorithm~\ref{alg:add-mem}, for each pair $(\timeitalic, \mem) \in \peak_{i-1}$, where $\timeitalic \ge \sigma(n_i)$, we set $\peak_{i}[\timeitalic] = \peak_{i-1}[\timeitalic] + \initmem(n_i)$, where $\initmem(n_i) = \sum_{d \in \outdata(n_i)} w(d)$ is the memory of the output data of $n_i$.
We study cases for $t \in [M(\sigma, p)]$. In the first three cases, it holds $t < \sigma(n_i) + c(n_i, P_j)$, and the data whose consumers become all scheduled in iteration $i$ have $\lend_d \ge \sigma(n_i) + c(n_i, P_j) > t$, hence they remain in $A_i(t)$; that is, it holds $A_{i}(t) = A_{i-1}(t)$ if $t < \sigma(n_i)$, and $A_{i}(t) = A_{i-1}(t) \cup \outdata(n_i)$ otherwise.\\

\noindent \underline{Case $t < \sigma(n_i)$:} 
$$
\begin{array}{ll}
\sum_{d \in A_i(t)} w(d)&=\\[3pt]
\sum_{d \in A_{i-1}(t)} w(d)&\le\\[3pt]
\peak_{i-1}[\before_{i-1}(t)]&=\\[3pt]
\peak_{i}[\before_i(t)]
\end{array}
$$

\noindent \underline{Case $t = \sigma(n_i)$}: 
$$\begin{array}{ll}
 \sum_{d \in A_i(\sigma(n_i))} w(d)&=\\[3pt]
 \sum_{d \in A_{i-1}(\sigma(n_i))} w(d) + \initmem(n_i) &\le\\[3pt]
 \peak_{i-1}[\before_{i-1}(\sigma(n_i))] + \initmem(n_i) &=\\[3pt]
 \peak_i[\sigma(n_i)] &=\\[3pt]
 \peak_{i}[\before_{i}(\sigma(n_i))]
\end{array}
$$

\noindent \underline{Case $\sigma(n_i) < t < \sigma(n_i) + c(n_i, P_j)$}:
$$
\begin{array}{ll}
     \sum_{d \in A_i(t)} w(d)&=\\[3pt]
     \sum_{d \in A_{i-1}(t)} w(d) + \initmem(n_i) &\le\\[3pt]
     \peak_{i-1}[\before_{i-1}(t)] + \initmem(n_i) &= \\[3pt]
     \peak_{i}[\max\{\before_{i-1}(t), \sigma(n_i)\}] &=\\[3pt]
     \peak_{i}[\before_{i}(t)].
\end{array}
$$
The two last equalities follow since, if $\before_{i-1}(t) \le \sigma(n_i)$ is true, then it holds $\before_{i}(t) = \sigma(n_i)$ by Algorithm~\ref{alg:add-mem} (lines 2--7); otherwise, $\before_{i}(t) = \before_{i-1}(t)$.\\

\noindent \underline{Case $t\ge \sigma(n_i) + c(n_i, P_j)$:}
$$
\begin{array}{ll}
 \sum_{d \in A_i(t)} w(d) &=\\[3pt]
 \sum_{d \in A_{i-1}(t)} w(d) + \initmem(n_i) - \sum_{d' \in F_i(t)} w(d') &\le\\[3pt]
 \peak_{i-1}[\before_{i-1}(t)] + \initmem(n_i) - \sum_{d' \in F_i(t)} w(d') &= \\[3pt]
 \peak_i[\max\{\before_{i-1}(t), \sigma(n_i) + c(n_i, P_j)\}] &= \\[3pt]
 \peak_i[\before_{i}(t)]
\end{array}
$$
where $F_i(t) \defeq \{d \in D_{i-1} \;|\; n_i \in \Consumers(d) \subseteq N_i,\ \lend_d \le t\}$ denotes the data whose consumers become all scheduled in iteration $i$ and whose lifetime ends by $t$, i.e., the data subtracted at time $t$ by Algorithm~\ref{alg:sub-memory} (lines 6--8). Since, for $d \in F_i(t)$, $\lend_d$ is a time in $\peak_i$, it holds $\lend_d \le t$ if and only if $\lend_d \le \before_i(t)$. Hence, it holds $A_i(t) = (A_{i-1}(t) \setminus F_i(t)) \cup \outdata(n_i)$, and the last equalities follow from the updates in Algorithm~\ref{alg:add-mem} (line 9) and Algorithm~\ref{alg:sub-memory} (lines 6--8), since, if it holds $\before_{i-1}(t) < \sigma(n_i) + c(n_i, P_j)$, then we get $\before_i(t) = \sigma(n_i) + c(n_i, P_j)$, otherwise $\before_i(t) = \before_{i-1}(t)$.

To prove equality in case $i = |N|$, it suffices to show $\sum_{d \in \Alive(t)} w(d) \ge \peak_{|N|}[\before_{|N|}(t)]$ for all $t \in [M(\sigma,p)]$.
By Algorithms~\ref{alg:add-mem} and \ref{alg:sub-memory}, we get
$$
\peak_{|N|}[\before_{|N|}(t)] = \sum_{d \in D\;:\; \lbegin_d \le \before_{|N|}(t) \land \lend_d > t} w(d)
$$
and by restricting data begin times it follows 
$$
\begin{array}{ll}
\sum_{d \in D\;:\; \lbegin_d \le \before_{|N|}(t) \;\land\; \lend_d > t} w(d)& \le\\ \sum_{d \in D\;:\; \lbegin_d \le t \;\land\; \lend_d > t} w(d) & = \\\sum_{d \in \Alive(t)} w(d).
\end{array}
$$
\end{proof}

\begin{claim}\label{claim:respect-wmax}
Assume $\wmax \ge \max_{n\in N}{\initmem(n)}$. Then, Algorithm~\ref{alg:mem-aware-graph} returns a schedule $(\sigma, p) \in \Sigma_I^{\wmax}$ or None.
\end{claim}
\begin{proof}
If the algorithm returns a schedule, it holds $(\sigma, p) \in \Sigma_I$: by $\canstart(\cdot)$, each task starts not before its parents finish; by the idle windows, each PE executes one task at a time; and only PEs with $c(n, P_j) < \infty$ are examined.

It remains to show $W(\sigma, p) \le \wmax$. When a task $n$ is scheduled, the values of $\peak$ increase by $\initmem(n)$ from $\sigma(n)$ on, and decrease by at least $\minusmem(n)$ from $\sigma(n) + c(n, p(n))$ on, since the data whose only consumer is $n$ are released at its finish time and other released data only later. Algorithm~\ref{alg:memory-violated} checks these upper bounds on the new values of $\peak$, at the existing times and at the new times $\sigma(n)$ and $\sigma(n) + c(n, p(n))$, so all values of $\peak$ remain at most $\wmax$; in the first iteration, they are at most $\initmem(n) \le \wmax$ by our assumption on $\wmax$. By Claim~\ref{claim:no-underestimate-memory},
$$W(\sigma, p) = \max_{t \in [M(\sigma,p)]} \sum_{d \in \Alive(t)} w(d) = \max_{t \in [M(\sigma,p)]} \peak_{|N|}[\before_{|N|}(t)] \le \wmax.$$
\end{proof}

% Limitations
\paragraph{Limitation 1: Guarantee.} Claim~\ref{claim:respect-wmax} guarantees $(\sigma, p) \in\Sigma_{I}^{\wmax}$ when a schedule is returned, but there is no guarantee a schedule will be found even if $\Sigma_{I}^{\wmax} \neq \emptyset$. This is due to the greedy choices, without backtracking, and to the potential memory overestimation in Algorithms~\ref{alg:memory-violated} and \ref{alg:add-mem}, which prevent some schedules in $\Sigma_{I}^{\wmax}$ from being explored. For example, a single run may fail when the task sorting method starts a long chain of tasks producing large data early, while the memory-optimal order would finish another branch first. Instead of assuming data lifetimes originally extend until infinity, upper bounding the finish time of the last consumer of each data chunk, e.g., by using the critical path, may be a first workaround, which we leave for future work. Modifying the choice of $\sigma(\cdot)$ for each task from first-fit to last-fit might also help, but may lead to a worse makespan. Missing a schedule that exists is the price to pay for scalability; search and mathematical programming methods avoid it, at a much higher running time. In practice, this happens mostly under the tightest memory limits, see Section~\ref{sec:experiments-scientific-workflows}; otherwise, a portfolio of four runs finds a schedule for almost all the instances.

\paragraph{Limitation 2: Non-monotonicity.} Suppose we execute Algorithm~\ref{alg:mem-aware-graph} twice on the same input $I$, with limits $\wmax^1$ and $\wmax^2$, where $\wmax^1 <\wmax^2$, and it returns $(\sigma, p)_1$ and $(\sigma, p)_2$, respectively. It may happen that $M((\sigma, p)_1) < M((\sigma, p)_2)$, since an early memory violation for $\wmax^1$ will force the algorithm to return another schedule, which might provide an improved makespan. Hence, a more relaxed memory limit may lead to a longer makespan.

\paragraph{Memory allocation.} The schedule returned by our algorithm defines the lifetime of each data chunk. Depending on the setting, memory is then allocated statically or dynamically. Assigning data chunks to memory addresses such that the total storage is minimized is the \textit{Dynamic Storage Allocation} problem (SR2 in \cite{garey1979computers}); the peak memory $W(\sigma,p)$, i.e., the maximum of $\peak$ at the end of the run (Claim~\ref{claim:no-underestimate-memory}), is a lower bound on the storage needed \cite{lamprou2023safe}. In practice, one may set $\wmax$ slightly below the size of the shared memory, to leave some slack for the memory allocator and any other external memory needs.

\subsection{Task Sorting Methods}\label{sec:task-sorting}
% task sorting methods
Different task sorting methods, recall the parameter $\mathit{SORT}$ in Algorithm~\ref{alg:mem-aware-graph}, may lead to different schedules.
We employ a variety of criteria to sort the set of ready tasks, such as:

\begin{enumerate}[topsep=0pt,label=(\roman*)]
    \item Max/Min (average) cost,
    \item Max/Min number of children of different type,
    \item Max/Min number of parents of different type,
    \item Max/Min bottom level,
    \item Max/Min top level, 
    \item Max/Min sum of bottom and top level \cite{ahmad1996analysis},
    \item Max difference of bottom to top level \cite{ahmad1996analysis},
    \item Max weighted length \cite{shirazi1990analysis},
    \item Max/Min depth,
    \item Reverse post ordering,
    \item Min added memory consumption,
    \item Max depth, where ties are broken in favour of larger output data  \cite{li2023s}.
\end{enumerate}
In our experiments, a task has the same cost on all PEs allowed to execute it, which is the cost used in (i) and in the levels. The bottom level of a task is the maximum cost of a path from the task to a sink task and the top level is the maximum cost of a path from a source task to the task. The depth of a task is the number of arcs of a longest path from a source task to it. The type of a task, e.g., computation or communication, is used in Section~\ref{sec:experiments-deep-learning} only. In (x), tasks are ordered by decreasing finish time in a depth-first search from the source tasks. In (xi), the key is $\initmem(n) - \minusmem(n)$. The keys of (i)--(x) and (xii) are precomputed in time linear in the size of the input; the key of (xi) uses $\minusmem(\cdot)$, which is accumulated during the run, see Algorithm~\ref{alg:sub-memory}, and is complete once the task is ready.

Besides the aforementioned task sorting methods, we introduce novel ones based on notions of topological sorting distance defined in \cite{antunovic2021evaluating}.
A topological sorting is a bijection $\pi$ from $N$ to $\{1, 2, \ldots, |N|\}$ such that for each $(n_1,n_2) \in A$ it holds $\pi(n_1) < \pi(n_2)$. 
Let $d_\pi(n_1,n_2) = \pi(n_2) - \pi(n_1)$ denote the \textit{distance} from $n_1$ to $n_2$ under topological sorting $\pi$. Then, for each task $n\in N$, the following distances from its parent tasks are defined:
%\begin{small}
\begin{itemize}
    \item[--] $s_\pi(n) \defeq \sum_{(n',n) \in A} d_\pi(n',n)$, the sum of the distances from the parents,
    \item[--] $\alpha_\pi(n)$ $\defeq 
    \left\{
    \begin{array}{ll}
        \frac{s_\pi(n)}{d^-(n)}, & \text{if } d^-(n) > 0, \\
        0, & \text{if } d^-(n) = 0,
    \end{array}
    \right.
    $ the average distance
    \item[--] $m_\pi(n) \defeq \max_{(n',n) \in A} d_\pi(n',n)$, the maximum distance,
\end{itemize}
%\end{small}
where $d^-(n)$ denotes the in-degree of $n$.

Our idea is to keep track of the number of tasks, say $i$, scheduled thus far in Algorithm~\ref{alg:mem-aware-graph}. Then, at iteration $i+1$, we assume $\pi(n) = i+1$ for each ready task $n$ and pick for scheduling among them the task $n$ which maximizes, for example, $s_\pi(n)$. We therefore define three new sorting methods (xiii) Max $s_\pi$, (xiv) Max $\alpha_\pi$, (xv) Max $m_\pi$.

We further define weight-sensitive versions, where $w(n_1,n_2) \defeq \sum_{d \in \alpha((n_1,n_2))} w(d)$ is the data weight shipped from $n_1$ to $n_2$:
\begin{itemize}
    \item[--] $d^w_\pi(n_1, n_2) = w(n_1, n_2) \cdot d_\pi(n_1, n_2)$, 
     \item[--] $s^w_\pi(n) =\sum_{(n',n) \in A} d^w_\pi(n',n)$,
    \item[--] $\alpha^w_\pi(n)$ $=
    \left\{
    \begin{array}{ll}
        \frac{s^w_\pi(n)}{W^-(n)}, & \text{if } W^-(n) > 0, \\
        0, & \text{if } W^-(n) = 0,
    \end{array}
    \right.
    $
    \item[--] $m^w_\pi(n) = \max_{(n',n) \in A} d^w_\pi(n',n)$,
\end{itemize}
where $W^-(n) = \sum_{(n',n) \in A} w(n',n)$ is the total data weight received by $n$ from its parents.
We then suggest new task sorting methods (xvi) Max $s^w_\pi$, (xvii) Max $\alpha^w_\pi$, and (xviii) Max $m^w_\pi$.
Intuitively, $d_\pi(n',n)$ counts the iterations for which the data shipped from $n'$ to $n$ has been waiting in memory: Max $s_\pi$, Max $\alpha_\pi$ and Max $m_\pi$ favour the ready task whose input data have waited the longest in total, on average, or for the oldest one, respectively. The weighted variants weight each waiting time by the data weight: Max $s^w_\pi$ favours the ready task whose input data have occupied the most memory over time, Max $\alpha^w_\pi$ the one whose input data have waited the longest on average per unit of weight, and Max $m^w_\pi$ the one with the single input that has occupied the most memory, so that no data stays in memory for long.

In the end, we keep only (xvi) Max $s^w_\pi$, which we call \ws. In our experiments, the unweighted variants (Max $s_\pi$, Max $\alpha_\pi$, Max $m_\pi$) and the other weighted variants (Max $\alpha^w_\pi$, Max $m^w_\pi$) are near-duplicates of \ws~or of each other, therefore we omit them: without them, the best schedule over all methods becomes worse on 1 of the 341 task graphs of Section~\ref{sec:experiments-scientific-workflows}, on a single PE, and only by $0.02\%$.

Note that it is not necessary to recompute the values from scratch at each iteration, since $\pi(n)$ increases by one for each candidate task $n$.
If we consider the $(i+1)$-th iteration, it holds $d^{(i+1)}_\pi(n',n) = (i+1) - \pi(n') = d^{(i)}_\pi(n',n) + 1$, and it follows
$s^{(i+1)}_\pi(n) = s^{(i)}_\pi(n) + d^-(n)$.
For $s^w$, the update adds $W^-(n)$, the total data weight received by $n$ from its parents. Hence, $s^w_\pi$ can be written in closed form as $s^w_\pi(n) = (i+1)\,W^-(n) - \sum_{(n',n) \in A} w(n',n)\,\pi(n')$, where the second term is fixed once $n$ becomes ready. In our \texttt{Python} implementation, the second term is computed once per ready task, which makes \ws~$3.5$ times faster on a single PE than recomputing the score at each iteration, with identical schedules.

\paragraph{\wspp: breaking the ties of \ws.} The score $s^w_\pi$ is often tied. On the scientific workflows of Section~\ref{sec:experiments-scientific-workflows}, $42\%$ of the decisions of \ws, that is, of its picks among two or more ready tasks, are ties; about half of them are among tasks whose score is $0$, i.e., tasks without parents or whose input data have zero size. A tie is then decided by the (arbitrary) task identifier, that is, by the order in which tasks appear in the input file. We define (xix) \wspp, which sorts by $s^w$ and breaks its ties by the key $u(n)$, see below, preferring larger values; remaining ties are broken by task identifier. The key is accumulated during the run: whenever a task $n'$ is scheduled, for each child $n''$ of $n'$ with $k \ge 1$ other parents not scheduled yet, the data $w(n',n'')$ starts waiting in memory for $n''$, and we add $\lfloor w(n',n'')/k \rfloor$ to $u(\cdot)$ of each of these $k$ parents. That is,
$$u(n) = \sum_{(n,n'') \in A}\ \sum_{\substack{(n',n'') \in A\\ n' \text{ scheduled}}} \left\lfloor \frac{w(n',n'')}{k_{n'}(n'')} \right\rfloor,$$
where $k_{n'}(n'')$ is the number of parents of $n''$ not scheduled right after $n'$ is scheduled. In words, for each child of $n$, we count the data already in memory waiting for that child, divided, when it starts waiting, among the parents of the child which are not scheduled yet. Scheduling $n$ brings such a child closer to being ready, and its input data closer to being released. Since $u(\cdot)$ only breaks ties of $s^w$, every decision where $s^w$ is not tied is the same for \wspp~and \ws. Both keys depend on the schedule so far, hence the ready tasks are re-ranked at each iteration. Besides the memory checks, this is a second adaptive element of our list scheduling: the sorting keys depend on the schedule so far. Updating $u(\cdot)$ costs $O\big(\sum_{n'' \,:\, (n',n'') \in A} d^-(n'')\big)$ per scheduled task $n'$.

We also tried ten other tie-breaking keys, e.g., the memory freed by $n$ or the size of the output data of $n$. None was clearly better than $u(\cdot)$ on the scientific workflows of Section~\ref{sec:experiments-scientific-workflows}.

\paragraph{Methods used in the experiments.} In Section~\ref{sec:experiments-scientific-workflows}, we refer to the methods by the following names: CostMax and CostMin (i), BottomLevelMax and BottomLevelMin (iv), TopLevelMax and TopLevelMin (v), BottomTopLevelMaxSum and BottomTopLevelMinSum (vi), BottomTopLevelComposite (vii), WeightedLength (viii), DepthMax and DepthMin (ix), ReversePostOrder (x), GreedyHeight (xi), DepthMaxMem (xii), \ws~(xvi), and \wspp~(xix). We do not use (ii) and (iii) there, since all tasks of a scientific workflow have the same type (homogeneous case), so their keys are constant. With the two PE rules of Section~\ref{sec:algo}, namely Load and ValidStart, the 17 methods give 34 pairs. In Section~\ref{sec:experiments-deep-learning}, tasks have a computation or a communication type, hence we also use methods based on task types, such as (ii) and (iii).

% algorithmic complexity
\subsection{Running Time}\label{sec:complexity}
The running time of Algorithm~\ref{alg:mem-aware-graph} depends on how many candidate placements are rejected because of the memory limit. We first bound the worst case and then the case without rejected placements.

Let $a = |A|$ and $m = |P|$. In this subsection only, $n$ denotes the number of tasks, i.e., $n = |N|$, and not a task as in the rest of the paper. The profile $\peak$ has at most $2n$ points, i.e., time-to-memory pairs $(t, \mem)$, since each scheduled task adds at most its start and finish time. Let $K$ denote the cost of a query of the form ``the maximum of $\peak$ over a range of times'' and of an update of the form ``add a value to $\peak$ over a range of times'', as well as of finding $\before(t)$ and of inserting a point. We discuss the value of $K$ after the complexity analysis claims below.

\begin{claim}\label{claim:running-time}
Assume the values by which the task sorting method ranks the ready tasks, namely its \textit{keys}, can be computed for all ready tasks in $O(n + a)$ time per iteration. Then, in the worst case, Algorithm~\ref{alg:mem-aware-graph} with the Load or ValidStart PE rule runs in $O\big(n^2 (n+m) K + |D|(n+K) + n\, m \log m\big)$ time. In the typical $m \le n$ case, Algorithm~\ref{alg:mem-aware-graph} runs in $O\big(n^3 K + |D|(n+K)\big)$ time.
\end{claim}
\begin{proof}
\textit{Initialization.} $\mathit{Initialize}(\cdot)$ visits each task, each arc, each input data chunk of each task, and each PE, i.e., $O(n + a + m + |D|\,n)$.

\textit{Iterations.} Each iteration schedules a task or returns None, so they are at most $n$.

\textit{Sizes.} At any time, there are at most $n$ ready tasks and at most $2n$ points in $\peak$. Each PE starts with one idle window, and each placement replaces one idle window by at most two (Algorithm~\ref{alg:update-idle}), hence there are at most $n + m$ idle windows over all PEs.

\textit{Sorting.} In an iteration, computing the keys costs $O(n+a)$ by assumption, sorting the ready tasks $O(n \log n)$, and sorting the PEs by load $O(m \log m)$, once per iteration, since loads do not change within $\mathit{ScheduleTask}(\cdot)$. Over all iterations, this is $O(n^2 \log n + n\,a + n\,m \log m)$.

\textit{Placement.} In an iteration, $\mathit{ScheduleTask}(\cdot)$ examines at most $n$ ready tasks, each against at most $n+m$ idle windows. For each pair of a task and an idle window, checking the time costs $O(1)$, and $\mathit{MemoryViolated}(\cdot)$ performs two range maxima and two lookups of $\before(\cdot)$, i.e., $O(K)$. $\mathit{UpdateIdle}(\cdot)$ costs $O(n+m)$. Over all iterations, this is $O(n^2(n+m)K)$.

\textit{Memory updates.} $\mathit{AddMemory}(\cdot)$ performs one range update and at most two insertions per iteration, i.e., $O(nK)$ in total. $\mathit{SubtractMemory}(\cdot)$ visits each input and output data chunk of the scheduled task, i.e., $O(|D|\,n)$ in total, since each data chunk has one producer and at most $n$ consumers, and performs a range update for a data chunk only when its last consumer is scheduled, i.e., at most once per data chunk, $O(|D|K)$ in total.

\textit{Ready tasks.} $\mathit{UpdateReady}(\cdot)$ visits each arc once, i.e., $O(a)$ in total.

\textit{Sum.} The total is $O(n^2(n+m)K + |D|(n+K) + n^2\log n + n\,a + n\,m\log m)$. Since we have $a \le n^2$ and $K \ge 1$, it holds $n^2 \log n + n\,a = O(n^3) \subseteq O(n^2(n+m)K)$, which gives the claimed bound. If $m \le n$, then $n\,m\log m = O(n^2 \log n)$ and $n^2(n+m)K = O(n^3K)$, so the total is $O(n^3K + |D|(n+K))$.
\end{proof}

If each arc carries $O(1)$ data chunks, then $|D| = O(a) \subseteq O(n^2)$, so $|D|(n+K) = O(n^3 + n^2K)$ and the bound becomes $O\big(n^2(n+m)K + n\,m\log m\big)$. This is a worst-case bound. If no placement is rejected because of the memory limit, the running time is much lower, see below.

Note that the assumption of Claim~\ref{claim:running-time} holds for our methods: static keys are precomputed once, in time linear in the size of the input (Section~\ref{sec:task-sorting}); the key of (xi) costs $O(1)$ per ready task; and the dynamic keys of \ws~and \wspp~cost $O(1)$ per ready task and $O(a)$ per iteration.

The bound in Claim~\ref{claim:running-time} counts every rejected placement. In the worst case, an iteration tries every ready task on every idle window, with a memory check each time. If the memory limit rejects no placement, an iteration schedules the first ready task, and on each PE it examines, only the first idle window where the task fits in time is checked for memory. This is one memory check per iteration with the Load rule and at most $m$ with ValidStart.

\begin{claim}\label{claim:running-time-no-rejection}
Under the assumption of Claim~\ref{claim:running-time}, if no call of $\mathit{MemoryViolated}(\cdot)$ returns True, Algorithm~\ref{alg:mem-aware-graph} runs in $O\big(n^2\log n + n\,a + n\,m\log m + |D|(n+K) + nK\big)$ time with the Load PE rule and in $O\big(n^2\log n + n\,a + |D|(n+K) + n\,m\,K\big)$ time with ValidStart.
\end{claim}
\begin{proof}
As for Claim~\ref{claim:running-time}, except for the placement. ValidStart does not sort the PEs by load, so the term $n\,m\log m$ is absent. In each iteration, finding the first idle window where the task fits in time costs $O(n+m)$, and there is one memory check with Load and at most $m$ with ValidStart, i.e., $O(nK)$ and $O(n\,m\,K)$ in total.
\end{proof}

For $|D| = O(a)$ and $K = O(n)$, the bound for Load is $O(n^2\log n + n\,a + n\,m\log m)$, i.e., the cost of computing the keys and sorting.

\paragraph{The value of $K$.} If $\peak$ is kept in an ordered map and ranges are scanned, as written in Algorithms~\ref{alg:memory-violated}, \ref{alg:add-mem}, and \ref{alg:sub-memory}, then $K = O(n)$. If its points are kept in consecutive blocks of $B$ points, each block storing the maximum memory of its points and an offset pending to be added to all of them, then $K = O(n/B + B)$, i.e., $K = O(\sqrt{n})$ for $B = \Theta(\sqrt{n})$, and with a balanced search tree storing the same per subtree, $K = O(\log n)$. For $m \le n$ and $|D| = O(n^2)$, the bound of Claim~\ref{claim:running-time} is then $O(n^4)$, $O(n^{3.5})$, and $O(n^3 \log n)$, respectively. We use blocks, see below.

\paragraph{A faster memory profile.} Algorithms~\ref{alg:memory-violated}, \ref{alg:add-mem}, and \ref{alg:sub-memory} access $\peak$ only through four operations: the maximum over a range of times (possibly a single time), adding a value over a range of times, finding $\before(t)$, and inserting a point. In our \texttt{Python} implementation, we keep the points of $\peak$ in blocks of $B = 128$ consecutive points; a block is split in two when it has more than $2B$ points. Each block stores its points, the maximum of their values, and an offset, which is pending to be added to all values of the block. The structure stores the same points as the ordered map, the value of a point being its stored value plus the offset of its block, and, since memory sizes are integers, all sums are exact. Hence, each operation returns the same answer with both representations, and the algorithm makes the same decisions. Since $B$ is fixed, $K = O(n)$ asymptotically. We also checked the equivalence in practice, on all our task graphs, where both representations gave identical schedules, and on 2,000,000 random operations of the four types above, applied to both representations (naive and block), which gave the same answers. On a montage graph where a run fails only after scheduling most tasks, the running time drops from $451.6$ to $7.75$ seconds.

\section{Experimental Results}\label{sec:experiments}
We evaluate our algorithm in two settings. In Section~\ref{sec:experiments-scientific-workflows}, we use a \texttt{Python} implementation of our algorithm to compare with MASTAG \cite{fradet2025parallel}, which is also implemented in \texttt{Python}, on 341 task graphs of scientific workflows. There, as in \cite{fradet2025parallel}, schedules are computed and evaluated, but not executed. In Section~\ref{sec:experiments-deep-learning}, we use our \texttt{C++} implementation of our algorithm inside MindSpore to schedule training graphs of large language models, and we measure the real training step time on Ascend accelerators. There, we also compare the \texttt{Python} implementation with the sequential solver of MASTAG on 7 training graphs with up to 115,528 tasks to showcase scalability. In Section~\ref{sec:experiments-discussion}, we summarize the comparison and discuss threats to its validity.

\subsection{Scientific Workflows}\label{sec:experiments-scientific-workflows}

\paragraph{Baseline.}
We experiment with memory-aware graph scheduling in the context of scientific workflows. A summary of the state-of-the-art approach of \citet{fradet2025parallel} follows. Initially, a memory-optimal sequential schedule is computed. Although this is NP-hard, see discussion just after Definition~\ref{def:graph-schedule-memory}, a worst-case quartic-complexity schedule graph compression algorithm, optionally followed by branch and bound, see \cite{fradet2023sequential,fradet2025graph}, is said to derive the desired sequential schedule with (near)-optimal memory for the task graphs considered in \cite{fradet2025parallel}.
Then, the goal is to minimize makespan under that tight, or a more relaxed, memory limit. In \cite{fradet2025parallel}, a  parallel schedule is derived by a novel algorithm using the current memory occupancy, the memory incurred by the new task to be scheduled, and the not-yet-scheduled part of the precomputed optimal-memory sequential schedule. The ready list of tasks is sorted by \textit{(i)} the optimal-memory sequential schedule, or \textit{(ii)} bottom levels, or \textit{(iii)} a mixing of the two. Unlike previous works \cite{marchal2018parallel, bathie2021dynamic}, the algorithm is guaranteed to return a schedule respecting the tight memory limit. 
Below, we call this approach \mastag, after the name of its public implementation \cite{mastagrepo}, and its sequential solver \lctes, after the venue of \cite{fradet2023sequential}. \mastag~derives three parallel schedules, one per ready list order \textit{(i)}--\textit{(iii)}. Its implementation reports all three, and we take the best one as its result.
Below, we compare our algorithm to \mastag~on more task graphs of scientific workflows than in \cite{fradet2025parallel}.

\paragraph{Benchmark.} We use 341 task graphs from 17 scientific workflow families, see Table~\ref{tab:benchmark}. They comprise the 126 Pegasus task graphs distributed with \mastag~(synth-genome, synth-ligo, and synth-montage, generated with the Pegasus workflow generator \cite{silva2014community}; 42 per application, 21 with about 50 tasks and 21 with about 100), and 215 task graphs of the WfCommons workflow instances \cite{coleman2022wfcommons}, recorded from Pegasus, Makeflow, Nextflow, Snakemake, and StreamFlow executions. All graphs are given in the DAX format read by \mastag. Task costs are execution times and data weights are file sizes, as given in the instances. A file that no task produces is in memory from the earliest start to the latest end of its consumers.

\paragraph{Setup.} We use the public implementation of \mastag~\cite{mastagrepo}, built with its own container recipe. We apply three small patches: a bug fix in the size of files used by a single task, and two fixes in time accounting and logging that do not change any schedule. Our algorithm is a \texttt{Python} port of our \texttt{C++} implementation, run with the library versions of the \mastag~container. All runs use an Intel Xeon E5-1660 v4 CPU (8 cores, 3.20GHz), one run at a time when timed. \lctes~uses a time limit of 600 seconds for its branch and bound, the default of the experiment configuration of \mastag, also used to reproduce the results of \cite{fradet2025parallel}. This limit applies to each call of the branch and bound, not to the whole run, so \lctes~may run for hours on a single graph. When the limit is reached, the best schedule found so far is returned. Since \mastag~does not always report such a time-out, we report both the graphs where \lctes~reports a time-out and the graphs where it runs for more than 600 seconds. We stop \mastag~after one day per task graph. On 4 and 8 PEs, we stop each of the three parallel schedulers of \mastag~after 600 seconds per call, as for each task sorting method of our algorithm, and our methods run one after the other on one thread. The schedulers (ii) and (iii) of \mastag~have an optional refinement, which updates the sequential schedule while the parallel schedule is built. It is on in the default experiment configuration of \mastag~and in the workflow experiments of \cite{fradet2025parallel}, on graphs of 50 and 100 tasks, but reported there as not practical for graphs with large ready lists. We use the refinement, unless stated otherwise. The running time of our algorithm is the time of the whole scheduler call, including its precomputations, e.g., of bottom levels, but not reading the input, as for the time reported by \mastag~for \lctes. For our four main comparisons with \mastag~(peak memory and running time on a single PE, and makespan and running time on 4 and 8 PEs), the ratios over task graphs have heavy tails, e.g., due to the time limit of \lctes. Hence, we give both their median and their geometric mean, computed for each limit and number of PEs; a range such as ``$0.2\%$ to $4.6\%$ on median'' goes from the smallest to the largest of these values. All other ratios aggregated over task graphs are geometric means.

\paragraph{Memory model.} Both tools compute memory in the same way: given the same task order, they return the same peak memory, except on 5 synth-ligo instances, where \mastag~merges two files with the same name, written by different tasks, into one (a difference of at most 50 KB). Hence, differences in memory between the tools come from different task orders.

\begin{table}[ht]
\centering
\footnotesize
\caption{Benchmark: number of task graphs, number of tasks, and running time in seconds on a single PE of \wspp~and of \lctes~(on the graphs where it returns a schedule); the \lctes~time limit 600 seconds applies to each call of its branch and bound, not to the whole run.}
\label{tab:benchmark}
\begin{tabular}{l|r|rrr|rr|rr}
\hline
\multirow{2}{*}{Workflow} & \multirow{2}{*}{Graphs} & \multicolumn{3}{c|}{Tasks} & \multicolumn{2}{c|}{\wspp~time (s)} & \multicolumn{2}{c}{\lctes~time (s)} \\
 & & min & median & max & median & max & median & max \\
\hline
1000genome & 22 & 52 & 387 & 902 & 0.058 & 0.244 & 609 & 623 \\
blast & 15 & 43 & 103 & 303 & 0.023 & 0.149 & 0.03 & 0.28 \\
bwa & 15 & 104 & 1,004 & 1,004 & 1.446 & 1.532 & 2.8 & 2.9 \\
cycles & 24 & 67 & 662 & 6,543 & 0.182 & 4.642 & 1,004 & 38,001 \\
epigenomics & 26 & 41 & 511 & 1,695 & 0.070 & 0.378 & 0.30 & 1.5 \\
montage & 17 & 58 & 748 & 9,805 & 0.157 & 14.882 & 3.9 & 1,082 \\
nextflow & 15 & 11 & 127 & 265 & 0.017 & 0.046 & 609 & 633 \\
synth-genome & 42 & 47 & 74 & 100 & 0.013 & 0.028 & 0.01 & 0.03 \\
synth-ligo & 42 & 50 & 75 & 100 & 0.013 & 0.034 & 2.0 & 281 \\
synth-montage & 42 & 50 & 75 & 100 & 0.016 & 0.025 & 0.09 & 0.16 \\
RASflow & 7 & 11 & 51 & 501 & 0.008 & 0.127 & 0.01 & 0.15 \\
rna-seq-star-deseq2 & 19 & 45 & 93 & 251 & 0.015 & 0.072 & 0.17 & 607 \\
seismology & 11 & 101 & 601 & 1,101 & 0.326 & 1.125 & 0.35 & 1.4 \\
soykb & 10 & 96 & 311 & 676 & 0.073 & 0.317 & 6.4 & 627 \\
srasearch & 25 & 22 & 64 & 104 & 0.012 & 0.019 & 0.04 & 0.07 \\
streamflow & 1 & 26 & 26 & 26 & 0.005 & 0.005 & 0.56 & 0.56 \\
varlociraptor & 8 & 91 & 307 & 1,783 & 0.040 & 0.373 & 9.9 & 110 \\
\hline
All & 341 & 11 & 100 & 9,805 & 0.018 & 14.882 & 0.23 & 38,001 \\
\hline
\end{tabular}
\end{table}

\paragraph{Comparison I: (Near)-optimal memory for a sequential schedule.}

We run Algorithm~\ref{alg:mem-aware-graph} on a single PE with $\wmax = \infty$, with each of the 17 task sorting methods of Section~\ref{sec:task-sorting}, and compare the peak memory to the one of the sequential schedule of \lctes. On a single PE and without a limit, no memory check is needed and each method returns one sequential order. To compare methods within our portfolio, we divide the peak memory of a method by the lowest peak among the 17 methods on the same task graph, and we take the geometric mean per workflow and then over the 17 workflows, so that each workflow counts the same. We call this value the \textit{score} of a method. A score of $1$ is best.

In Table~\ref{tab:onecore}, we present the results per workflow. \wspp~has score $1.048$, against $1.092$ for \ws, and its worst workflow improves from $2.349$ (RASflow) to $1.386$ (rna-seq-star-deseq2). The two methods give the same peak on 306 out of 341 task graphs. \wspp~is better on 11 and worse on 24, in five workflows (8 in nextflow). To check that the choice of \wspp~does not overfit our benchmark, we leave out one workflow at a time and select the method with the best score on the other 16 workflows: \wspp~is selected in 16 out of the 17 cases.

\begin{figure}[ht]
\centering
\begin{subfigure}[b]{0.49\linewidth}
\centering
\includegraphics[width=\linewidth]{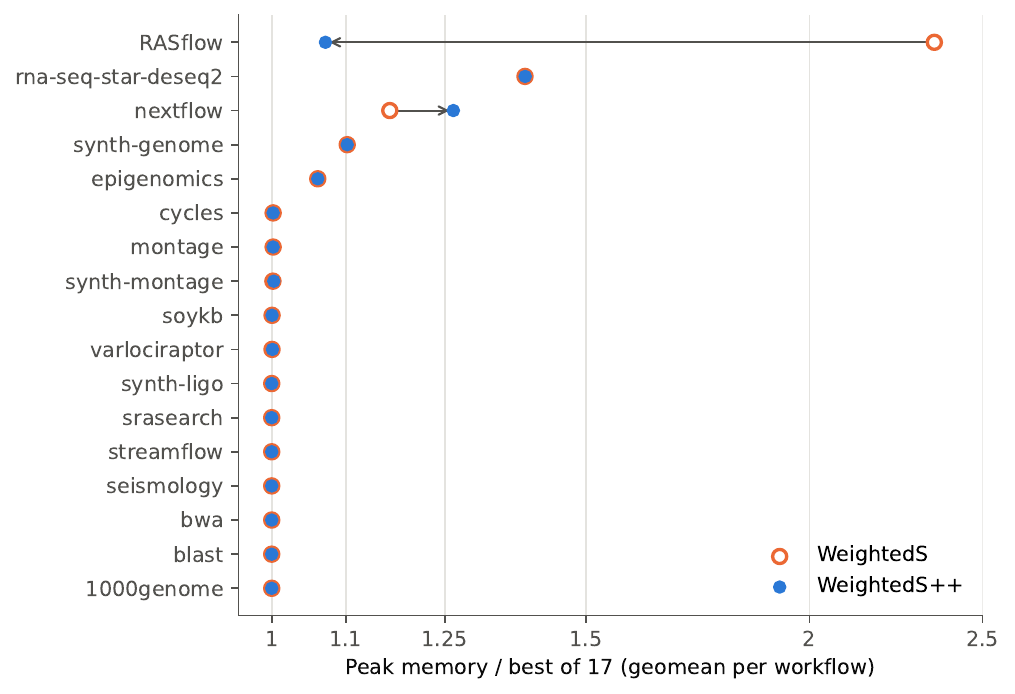}
\caption{Peak memory of \ws~and \wspp~relative to the best of our 17 methods, per workflow (geometric mean).}
\label{fig:onecore-per-workflow}
\end{subfigure}\hfill
\begin{subfigure}[b]{0.49\linewidth}
\centering
\includegraphics[width=\linewidth]{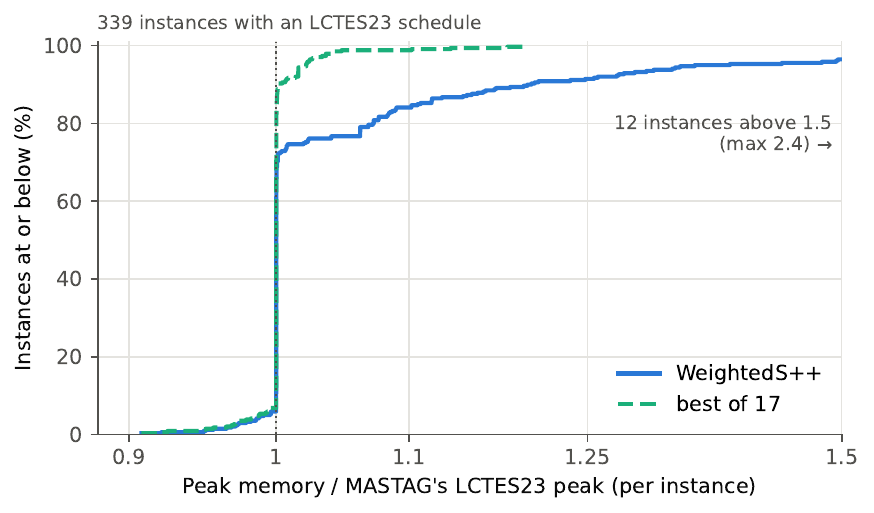}
\caption{Share of graphs with a peak at most $x$ times that of \lctes; left of $1$: lower peaks (cycles, where \lctes~times out).}
\label{fig:onecore-vs-mastag}
\end{subfigure}
\caption{Single PE, peak memory.}
\label{fig:onecore}
\end{figure}

Compared to \lctes, \wspp~has a median ratio below $1.0001$ and a geometric mean ratio of $1.050$ over the 339 task graphs where \lctes~returns a schedule. It finds exactly the same peak on $29\%$ of them, a peak within $0.1\%$ on $70\%$, and a lower peak on $5.9\%$, all in the cycles workflow, where \lctes~stops at its time limit. If we take the best of our 17 methods per task graph, the geometric mean ratio becomes $1.0015$, within $0.1\%$ on $89\%$ of the graphs, and lower on $6.8\%$. The remaining gap is concentrated in five workflows, nextflow, rna-seq-star-deseq2, synth-genome, RASflow, and epigenomics, where another method of the portfolio does better than \wspp, see Figures~\ref{fig:onecore-per-workflow} and \ref{fig:onecore-vs-mastag}.

\lctes~reports a time-out on 48 task graphs, mostly cycles (22) and 1000genome (18), and runs for more than 600 seconds on 57, these 48 and 9 more, among which 8 nextflow graphs. It returns no schedule within a day on 2 more cycles graphs. When the time limit is reached, the returned schedule is not proven optimal: on all 22 of these cycles graphs, the best of our 17 methods finds a lower peak, and \wspp~alone on 20 of them. Also, on one nextflow graph (hic-dirt02-001), \lctes~finishes within the limit, yet CostMax finds a $7.5\%$ lower peak, under \mastag's own memory evaluation. On the other hand, \wspp~runs in 18 milliseconds on median and in 15 seconds on the largest graph, see Table~\ref{tab:benchmark}. On the 339 task graphs where \lctes~returns a schedule, \lctes~takes 31.4 hours in total and \wspp~73 seconds. Per graph, \wspp~is $5.5$ times faster on median and $23.2$ times on geometric mean, and $713$ times on geometric mean on the 119 graphs where \lctes~takes more than one second. \wspp~is slower on 29 small graphs, where both tools take at most $0.12$ seconds. Running all 17 methods one after the other takes about 5 minutes in total.

\begin{table*}[ht]
\centering
\small
\caption{Single PE, peak memory. To best: ratio to the best of our 17 methods. Vs \lctes: ratio to \lctes, and share of graphs within $0.1\%$ of it or below, on the graphs where it returns a schedule (in parentheses, if fewer). Ratios are geometric means. Time-outs: graphs where \lctes~reports a time-out (in parentheses, if different: graphs where it runs for more than 600 seconds). In row All, the to-best columns are the score (geometric mean over workflows), and the other columns pool the 339 graphs.}
\label{tab:onecore}
\resizebox{\linewidth}{!}{
\begin{tabular}{l|r|cc|cc|cc|r}
\hline
\multirow{2}{*}{Workflow} & \multirow{2}{*}{Graphs} & \multicolumn{2}{c|}{To best} & \multicolumn{2}{c|}{\wspp~vs \lctes} & \multicolumn{2}{c|}{Best of 17 vs \lctes} & \multirow{2}{*}{Timeouts} \\
 & & \ws & \wspp & ratio & $\le 0.1\%$ & ratio & $\le 0.1\%$ & \\
\hline
1000genome & 22 & 1.000 & 1.000 & 1.000 & 100\% & 1.000 & 100\% & 18 \\
blast & 15 & 1.000 & 1.000 & 1.000 & 100\% & 1.000 & 100\% & 0 \\
bwa & 15 & 1.000 & 1.000 & 1.000 & 100\% & 1.000 & 100\% & 0 \\
cycles & 24 (22) & 1.002 & 1.002 & 0.974 & 91\% & 0.972 & 100\% & 22 \\
epigenomics & 26 & 1.061 & 1.061 & 1.066 & 15\% & 1.005 & 69\% & 0 \\
montage & 17 & 1.002 & 1.002 & 1.003 & 65\% & 1.001 & 88\% & 0 (1) \\
nextflow & 15 & 1.164 & 1.264 & 1.307 & 20\% & 1.034 & 53\% & 1 (9) \\
synth-genome & 42 & 1.102 & 1.102 & 1.103 & 50\% & 1.000 & 90\% & 0 \\
synth-ligo & 42 & 1.000 & 1.000 & 1.000 & 100\% & 1.000 & 100\% & 0 \\
synth-montage & 42 & 1.002 & 1.003 & 1.003 & 86\% & 1.000 & 100\% & 0 \\
RASflow & 7 & 2.349 & 1.072 & 1.072 & 0\% & 1.000 & 100\% & 0 \\
rna-seq-star-deseq2 & 19 & 1.386 & 1.386 & 1.421 & 0\% & 1.025 & 11\% & 3 \\
seismology & 11 & 1.000 & 1.000 & 1.000 & 100\% & 1.000 & 100\% & 0 \\
soykb & 10 & 1.000 & 1.000 & 1.000 & 100\% & 1.000 & 100\% & 4 \\
srasearch & 25 & 1.000 & 1.000 & 1.000 & 100\% & 1.000 & 100\% & 0 \\
streamflow & 1 & 1.000 & 1.000 & 1.000 & 100\% & 1.000 & 100\% & 0 \\
varlociraptor & 8 & 1.000 & 1.001 & 1.001 & 25\% & 1.000 & 100\% & 0 \\
\hline
All & 341 (339) & 1.092 & 1.048 & 1.050 & 70\% & 1.001 & 89\% & 48 (57) \\
\hline
\end{tabular}
}
\end{table*}

\begin{figure}[ht]
\centering
\includegraphics[width=0.5\linewidth]{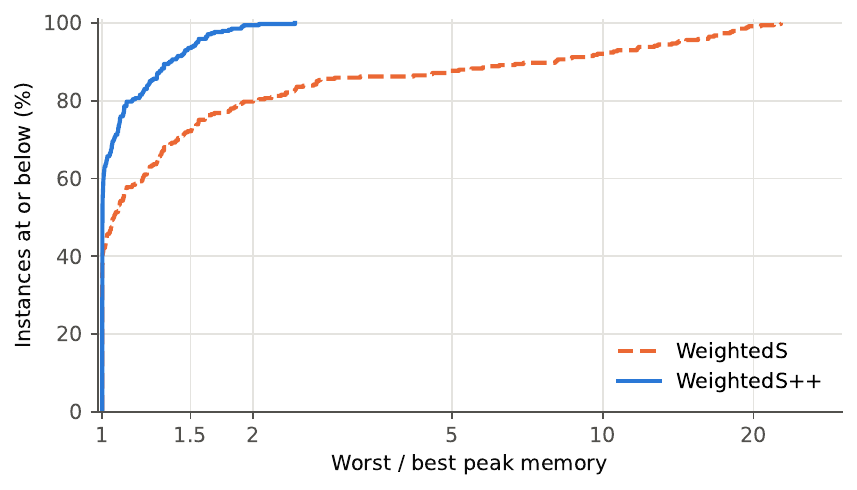}
\caption{Input-order sensitivity: share of task graphs where the ratio of the worst to the best peak memory over 11 task orders (file order and 10 random permutations) is at most $x$.}
\label{fig:order}
\end{figure}

 \paragraph{Input-order sensitivity.} Each sorting method breaks its remaining ties by task identifier, that is, by the order of tasks in the input file. To measure the effect of this order, we run each task graph with its file order and with 10 random permutations of the task identifiers (the graph remains the same). The peak of \ws~depends on the order for 236 out of 341 task graphs, and the ratio of the worst to the best peak over the 11 orders is $1.600$ on geometric mean and up to $22.7$ (e.g., 2.5 GB with the file order and 58 GB with every random order on one 1000genome graph). For \wspp, the same values are 230, $1.085$, and $2.4$, see Figure~\ref{fig:order}. For example, \ws~matches the best of our 17 methods on 1000genome and srasearch with the file order, but is $8.6$ and $5.0$ times the best, respectively, on geometric mean over the 10 random orders, whereas \wspp~is within $0.15\%$ of the best on every order. Hence, part of the good results of \ws~is due to a favourable input order, and the tie-break of \wspp~removes most of this dependence. Synth-genome, and parts of rna-seq-star-deseq2 and nextflow, are ranking problems and the tie-break of \wspp~does not improve them. The file order, which we do not choose, also favours \wspp, but much less than \ws: with random orders, the geometric mean ratio to \lctes~is $1.084$ instead of $1.050$ for \wspp, mostly on cycles and synth-ligo, and $1.490$ instead of $1.063$ for \ws.

\paragraph{Comparison II: Speeding up makespan under tight memory limits.}

We now consider 4 and 8 homogeneous PEs, like in \cite{fradet2025parallel}, and seven memory limits per task graph, see Table~\ref{tab:limits}. Limits of the M family interpolate between the peak of the sequential schedule of \lctes, which \mastag~computes as optimal unless \lctes~stops at its time limit, denoted $\mathit{minMem}$, and the peak of \mastag's parallel list schedule without a memory limit on the same number of PEs, denoted $\mathit{maxMem}$. M and M$+50\%$ are the two limits used in the experiments of \cite{fradet2025parallel}, as set in the script of \mastag~that reproduces them; we add M$+25\%$ and M$+75\%$. Limits of the G family relax the peak of the sequential schedule of \wspp. Limit M is the tightest, except on cycles, where \lctes~stops at its time limit. A schedule within M always exists, namely the sequential schedule of \lctes, but our algorithm has to find a schedule whose peak is at most this sequential peak. Note that the G limits are defined from our own sequential schedule, hence a schedule within them exists by construction (our sequential one), and our portfolio finds one on 4 and 8 PEs for all instances. For the G limits, we therefore report mainly the makespan. In total, we have 14 configurations (limit and number of PEs). The M limits are defined for the 339 task graphs where \lctes~returns a schedule.

\begin{table}[ht]
\centering
\small
\caption{Memory limits on 4 and 8 PEs}
\label{tab:limits}
\begin{tabular}{l|l}
\hline
Label & Memory limit \\
\hline
M & $\mathit{minMem}$: peak of the sequential schedule of \lctes \\
M$+c$ & $\mathit{minMem} + c\cdot(\mathit{maxMem} - \mathit{minMem})$, for $c \in \{25\%, 50\%, 75\%\}$ \\
G & peak of the sequential schedule of \wspp \\
G$+0.1\%$, G$+1\%$ & that peak $\times 1.001$, $\times 1.01$ \\
\hline
\end{tabular}
\end{table}

\paragraph{A portfolio of four runs.} With 17 task sorting methods and 2 PE rules, there are 34 pairs. Running all of them is possible, but we look for a few pairs which together schedule almost every instance with a short makespan. We run all 34 pairs on all instances and configurations, that is, 161,772 runs (only 28 of them, all on one montage graph, stop at our time limit of 600 seconds and count as infeasible), and we measure two quantities for any set of pairs, where the set returns its feasible schedule with the shortest makespan. Its \textit{feasibility} in a configuration is the share of instances it schedules within the limit, relative to the instances scheduled by at least one of the 34 pairs; its \textit{makespan} is the geometric mean ratio to the best of the 34 pairs, over the instances it schedules. We select four pairs:
\wspp/ValidStart, DepthMaxMem/ValidStart, BottomTopLevelMaxSum/Load, and BottomTopLevelComposite/ValidStart. A greedy selection, which adds the pair that improves the worst feasibility over the 14 configurations the most, with ties broken by the makespan, picks the same four pairs, with \ws~in place of \wspp. The two choices are equivalent (worst feasibility $0.982$ against $0.973$, mean feasibility $0.995$ against $0.996$, makespan $1.0306$ against $1.0302$), and we keep \wspp, the best method for peak memory on a single PE, whose peak depends much less on the order of the tasks in the input file than that of \ws.

In Table~\ref{tab:ensemble}, we present the results per memory limit. Under all limits but M, the portfolio schedules $99.7\%$ to $100\%$ of the instances, and its makespan is within $1.8\%$ to $5.1\%$ of the best of the 34 pairs. Under M, at least one pair finds a schedule for $64.9\%$ of the instances, and the portfolio for $63.4\%$ (4 PEs) and $63.1\%$ (8 PEs). That is, M is hard for our algorithm, since its peak may not exceed the sequential peak of \lctes, which \wspp~matches or beats on a single PE only for $35\%$ of the instances. On the other hand, \mastag~finds a schedule within M by construction, since it starts from its sequential schedule. 

\begin{figure}[ht]
\centering
\begin{subfigure}[b]{0.49\linewidth}
\centering
\includegraphics[width=\linewidth]{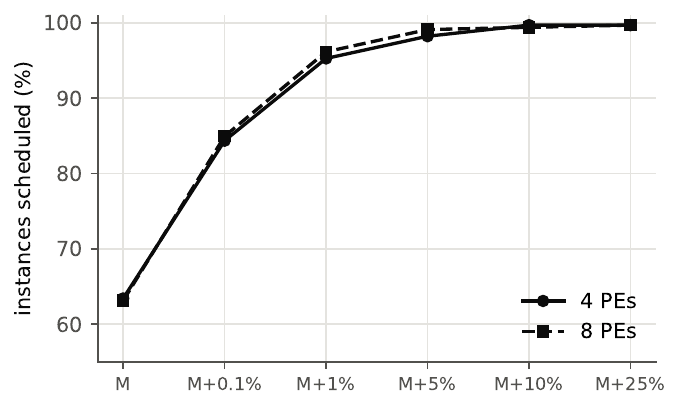}
\caption{Share of instances the portfolio schedules (\mastag~schedules all of them).}
\label{fig:mplus-feas}
\end{subfigure}\hfill
\begin{subfigure}[b]{0.49\linewidth}
\centering
\includegraphics[width=\linewidth]{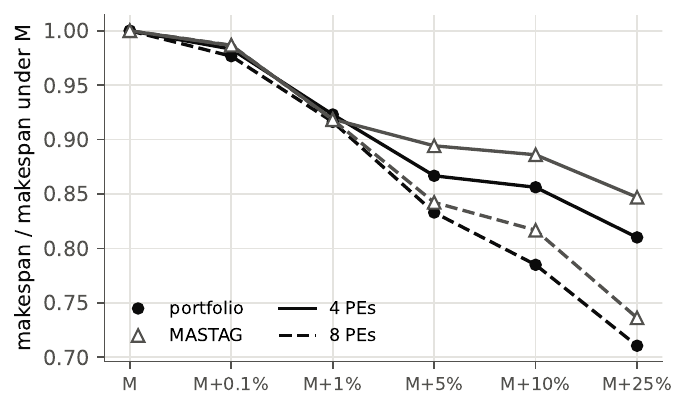}
\caption{Makespan of each tool, relative to its own makespan under M (geometric mean).}
\label{fig:mplus-mk}
\end{subfigure}
\caption{Limits slightly above M (M plus $0.1\%$ to $25\%$ of the gap between $\mathit{minMem}$ and $\mathit{maxMem}$). Each point in (b) is over the instances the portfolio schedules under both M and that limit. The x-axis is not to scale.}
\label{fig:mplus}
\end{figure}

Since M is hard for our algorithm, after the main runs we also run the portfolio and \mastag~under four limits between M and M$+25\%$, namely M$+0.1\%$, M$+1\%$, M$+5\%$, and M$+10\%$, with $c \in \{0.1\%, 1\%, 5\%, 10\%\}$ in the M family of Table~\ref{tab:limits}. The first two are very close to M: on median, they exceed M by only $0.03\%$ to $0.04\%$ and $0.35\%$ to $0.37\%$ of its value, respectively (for 4 and 8 PEs, as $\mathit{maxMem}$ depends on the number of PEs). Yet, the portfolio schedules $84\%$ to $85\%$ of the instances under M$+0.1\%$, $95\%$ to $96\%$ under M$+1\%$, $98\%$ to $99\%$ under M$+5\%$, and more than $99\%$ under M$+10\%$, against $63\%$ under M (Figure~\ref{fig:mplus-feas}). \mastag~schedules all of them. On the instances that the portfolio already schedules under M, the makespan of the portfolio stays within $1\%$ of that of \mastag~on median under each of these limits. The instances that become schedulable are harder: there, the makespan of the portfolio is $13\%$ to $27\%$ longer than that of \mastag~on median, since \mastag~starts from the sequential schedule of \lctes. Over all instances both tools schedule, the makespan of the portfolio is $2.1\%$ to $7.1\%$ longer on median ($8.4\%$ to $17.5\%$ on geometric mean), and the portfolio is $7.5$ to $10.3$ times faster than \mastag~on median. Without the refinement of \mastag, the makespan of the portfolio is $2.1\%$ to $8.1\%$ longer on median. The extra memory also shortens the schedules of the portfolio: on the instances scheduled under both limits, its makespan under M$+1\%$ is $7.7\%$ (4 PEs) and $8.4\%$ (8 PEs) shorter than under M, on geometric mean, and $19\%$ and $29\%$ shorter under M$+25\%$, while that of \mastag~is $15\%$ and $26\%$ shorter on the same instances (Figure~\ref{fig:mplus-mk}). These four limits are not part of the other tables. 

On 55 (4 PEs) and 37 (8 PEs) of the 339 instances, $\mathit{maxMem} = \mathit{minMem}$, so all M limits coincide. Such trivial instances are excluded from the success rates of \cite{fradet2025parallel}, but not from its average speedups over all samples, and we include them everywhere. Without them, the portfolio schedules $56\%$ (4 PEs) and $59\%$ (8 PEs) of the instances under M, $81\%$ and $83\%$ under M$+0.1\%$, $94\%$ and $96\%$ under M$+1\%$, and $98\%$ and $99\%$ under M$+5\%$.

\begin{table}[ht]
\centering
\small
\caption{Portfolio of four pairs: share of instances scheduled within the limit and makespan (geometric mean ratio to the best of the 34 pairs, over the instances the portfolio schedules)}
\label{tab:ensemble}
%\resizebox{\columnwidth}{!}{
\begin{tabular}{l|r|r|r|cc}
\hline
\multirow{2}{*}{Limit} & \multirow{2}{*}{Graphs} & Some pair & Portfolio & \multicolumn{2}{c}{Makespan} \\
 & & schedules & schedules (4 / 8 PEs) & 4 PEs & 8 PEs \\
\hline
M & 339 & 64.9\% & 63.4\% / 63.1\% & 1.015 & 1.018 \\
M$+25\%$ & 339 & 100\% & 99.7\% / 99.7\% & 1.024 & 1.046 \\
M$+50\%$ & 339 & 100\% & 100\% / 100\% & 1.018 & 1.038 \\
M$+75\%$ & 339 & 100\% & 100\% / 100\% & 1.018 & 1.035 \\
G & 341 & 100\% & 100\% / 100\% & 1.023 & 1.032 \\
G$+0.1\%$ & 341 & 100\% & 100\% / 100\% & 1.044 & 1.051 \\
G$+1\%$ & 341 & 100\% & 100\% / 100\% & 1.025 & 1.039 \\
\hline
\end{tabular}
%}
\end{table}

\begin{figure}[ht]
\centering
\includegraphics[width=0.85\linewidth]{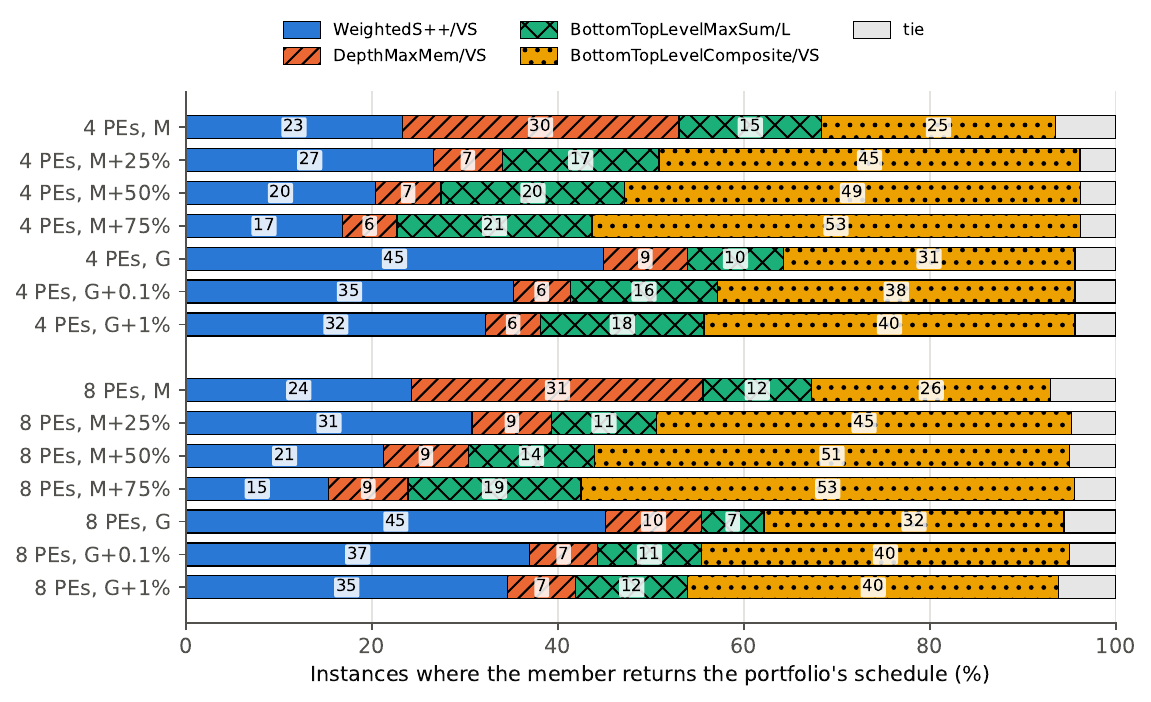}
\caption{Share of the instances where each member returns the best schedule of the portfolio.}
\label{fig:winners}
\end{figure}

Table~\ref{tab:buildup} shows the portfolio when we add its members one at a time, and when we remove one member from the full portfolio. With \wspp/ValidStart alone, the worst feasibility over the 14 configurations is $0.586$, mostly under M, and the makespan is $1.090$. With all four members, they are $0.973$ and $1.030$. Removing one member lowers the worst feasibility by $0.02$ to $0.24$ and increases the makespan by $0.4\%$ to $2.5\%$. Without \wspp, the mean feasibility drops the most, to $0.883$, mostly under the G limits. Without DepthMaxMem, the worst feasibility drops the most, to $0.732$, under M. A fifth member would add little: the best one, TopLevelMax/ValidStart, gives a worst feasibility of $0.982$ and a makespan of $1.026$. Finally, each member alone returns the best schedule of the portfolio on a large share of the instances (Figure~\ref{fig:winners}): \wspp~on $15\%$ to $45\%$ (most under G), DepthMaxMem on $6\%$ to $31\%$ (most under M), BottomTopLevelMaxSum on $7\%$ to $21\%$, and BottomTopLevelComposite on $25\%$ to $53\%$. Ties ($4\%$ to $7\%$ of the instances) are not counted.

\begin{table}[ht]
\centering
\small
\caption{The portfolio when members are added one at a time (top) and when one member is removed from the four (bottom). Feasibility is the share of instances scheduled, among those that at least one of the 34 pairs schedules, worst and mean over the 14 configurations. Makespan is the ratio to the best of the 34 pairs (geometric mean).}
\label{tab:buildup}
\begin{tabular}{l|ccc}
\hline
 & Worst feas. & Mean feas. & Makespan \\
\hline
\multicolumn{4}{l}{\emph{Adding members one at a time}} \\
\wspp/ValidStart & 0.586 & 0.898 & 1.090 \\
+ DepthMaxMem/ValidStart & 0.868 & 0.949 & 1.071 \\
+ BottomTopLevelMaxSum/Load & 0.950 & 0.978 & 1.052 \\
+ BottomTopLevelComposite/ValidStart & 0.973 & 0.996 & 1.030 \\
(+ TopLevelMax/ValidStart, not used) & 0.982 & 0.998 & 1.026 \\
\hline
\multicolumn{4}{l}{\emph{Full portfolio without one member}} \\
without \wspp/ValidStart & 0.742 & 0.883 & 1.056 \\
without DepthMaxMem/ValidStart & 0.732 & 0.960 & 1.034 \\
without BottomTopLevelMaxSum/Load & 0.918 & 0.976 & 1.037 \\
without BottomTopLevelComposite/ValidStart & 0.950 & 0.978 & 1.052 \\
\hline
\end{tabular}
\end{table}

Since we select the portfolio on the instances we report, we check for overfitting: we repeat the greedy selection on 16 workflows and evaluate the selected pairs on the remaining one, for each of the 17 workflows. On the held-out workflows, the worst feasibility is $0.941$, against $0.982$ for the same selection on all workflows and $0.973$ for our portfolio, and the makespan $1.035$, against $1.031$ and $1.030$. In 14 of the 17 folds, the selection picks the same four task sorting methods, up to the choice between \ws~and \wspp, and 13 of these folds pick \ws.

In Table~\ref{tab:speedup}, we present the speedup of the portfolio over the sequential schedule, i.e., the sum of task costs divided by the makespan, on the instances it schedules. Along the M family, the speedup grows with the limit, up to $3.34$x on 4 PEs and $5.58$x on 8 PEs. Under G, where every instance is scheduled, $1\%$ more memory increases the speedup by $13\%$ on 4 PEs and by $19\%$ on 8 PEs.

\begin{table}[ht]
\centering
\small
\caption{Speedup of the portfolio over the sequential schedule (geometric mean over the instances it schedules; 215 and 214 instances under M for 4 and 8 PEs, respectively)}
\label{tab:speedup}
\begin{tabular}{l|cc}
\hline
Limit & 4 PEs & 8 PEs \\
\hline
M & 2.34 & 3.23 \\
M$+25\%$ & 3.05 & 4.79 \\
M$+50\%$ & 3.24 & 5.28 \\
M$+75\%$ & 3.34 & 5.58 \\
G & 2.41 & 3.31 \\
G$+0.1\%$ & 2.60 & 3.63 \\
G$+1\%$ & 2.72 & 3.95 \\
\hline
\end{tabular}
\end{table}

We say the memory limit changes a placement when it rejects the placement that the task sorting method and the PE rule would choose without a limit. For the schedule returned by the portfolio, the limit changes at least one placement in $56\%$ to $86\%$ of the instances, depending on the configuration, and $4\%$ to $21\%$ of all placements (260k out of 1.75M). It changes as many or more placements on 8 PEs than on 4. Within the M family, it changes more under tighter limits, $15\%$ to $16\%$ under M against $4\%$ to $7\%$ under M$+75\%$. It changes $18\%$ to $21\%$ of the placements under the G limits. Most often, the task moves to another PE. Less often, the next ready task is chosen instead, or the task moves to a later idle window of the same PE. The distribution is skewed: on median, an instance has $0.3\%$ to $9\%$ of its placements changed, while one instance in ten has $16\%$ to $60\%$. Per member, over all its feasible runs, \wspp/ValidStart almost never overrides its own priority order ($0.08\%$ of placements) and adapts to the limit by moving tasks in time or to other PEs, while the other three members choose a different task in $5.0\%$ to $9.3\%$ of their placements. We give the full numbers in Table~\ref{tab:binding}.

\begin{table}[ht]
\centering
\small
\caption{Placements changed by the memory limit, for the schedules returned by the portfolio: share of instances with at least one change, share of all placements changed, split by where the task went instead, and median and 90th percentile per instance. Among members with the same makespan, the portfolio returns the one with the lower peak memory.}
\label{tab:binding}
%\resizebox{\columnwidth}{!}{
\begin{tabular}{l|c|c|ccc|cc}
\hline
\multirow{2}{*}{Configuration} & Instances with & Placements & Later & Other & Different & \multicolumn{2}{c}{Per instance} \\
 & a change & changed & window & PE & task & median & 90th pct \\
\hline
4 PEs, M & 74\% & 15.4\% & 0.7\% & 11.9\% & 2.8\% & 3.1\% & 51\% \\
4 PEs, M$+25\%$ & 66\% & 14.0\% & 2.2\% & 7.4\% & 4.4\% & 3.0\% & 47\% \\
4 PEs, M$+50\%$ & 60\% & 9.0\% & 1.4\% & 6.2\% & 1.5\% & 1.2\% & 25\% \\
4 PEs, M$+75\%$ & 56\% & 3.7\% & 0.3\% & 2.7\% & 0.7\% & 0.3\% & 16\% \\
4 PEs, G & 76\% & 18.1\% & 1.8\% & 15.0\% & 1.3\% & 6.0\% & 49\% \\
4 PEs, G$+0.1\%$ & 65\% & 18.0\% & 1.4\% & 15.0\% & 1.5\% & 5.1\% & 50\% \\
4 PEs, G$+1\%$ & 57\% & 20.4\% & 1.4\% & 14.7\% & 4.3\% & 2.9\% & 60\% \\
8 PEs, M & 83\% & 16.4\% & 0.7\% & 13.1\% & 2.6\% & 5.8\% & 53\% \\
8 PEs, M$+25\%$ & 81\% & 14.0\% & 1.8\% & 8.8\% & 3.3\% & 7.0\% & 41\% \\
8 PEs, M$+50\%$ & 73\% & 9.9\% & 1.8\% & 7.9\% & 0.2\% & 3.0\% & 25\% \\
8 PEs, M$+75\%$ & 69\% & 6.6\% & 1.6\% & 4.7\% & 0.2\% & 1.9\% & 18\% \\
8 PEs, G & 86\% & 19.9\% & 1.5\% & 16.2\% & 2.2\% & 9.3\% & 49\% \\
8 PEs, G$+0.1\%$ & 76\% & 19.6\% & 1.0\% & 16.4\% & 2.3\% & 9.0\% & 57\% \\
8 PEs, G$+1\%$ & 70\% & 21.2\% & 0.9\% & 16.1\% & 4.2\% & 8.4\% & 60\% \\
\hline
\end{tabular}
%}
\end{table}

\paragraph{Comparison to \mastag.} In Table~\ref{tab:vs-mastag-multi}, we compare the portfolio to \mastag~on 4 and 8 PEs. The result of \mastag~is its feasible schedule with the shortest makespan among its three parallel schedulers, ties broken by running time, and its running time is the time of \lctes~plus the time of its three schedulers. For the 2 cycles graphs where \lctes~returns no schedule, the M limits are not defined, since they need its peak, so we run them only under the G limits.

Under the M limits, \mastag~always finds a schedule, since it starts from its sequential schedule, whereas the portfolio finds one for $63.1\%$ to $63.4\%$ of the instances under M, and for $99.7\%$ to $100\%$ under M$+25\%$ to M$+75\%$. Under the G limits, the portfolio finds a schedule for all instances, and \mastag~for $93.5\%$ to $94.7\%$: it returns none on the 2 cycles graphs without an \lctes~schedule, and on 20 cycles graphs (16 under G$+1\%$) where the limit, derived from our sequential peak, is below the peak of \lctes, which stops at its time limit there. On the instances both tools schedule, the makespan of the portfolio is within $0.2\%$ to $4.6\%$ of that of \mastag~on median, and within $1.7\%$ to $17.0\%$ on geometric mean, although \mastag~is shorter on $77\%$ to $86\%$ of them. Under M, where the portfolio schedules $63\%$ of the instances, its makespan is within $0.2\%$ (4 PEs) and $0.8\%$ (8 PEs) of that of \mastag~on median. The larger gaps come from a few workflows, mainly montage, srasearch, and 1000genome. On 8 of the 17 workflows, the two tools are within $2.2\%$ on geometric mean over both numbers of PEs, and on 13 of the 17 they are within $10\%$ on each of 4 and 8 PEs. Table~\ref{tab:vs-mastag-workflow} gives the results per workflow.

The portfolio is $18$ to $30$ times faster than \mastag~(refinement on) on geometric mean, and $7$ to $12$ times on median. Without \lctes, the parallel schedulers of \mastag~alone are $3.9$ to $10.3$ times slower than the portfolio on geometric mean. In total, the portfolio takes 3.3 hours over all configurations, against 47.9 hours for the parallel schedulers of \mastag~and 33.0 hours for \lctes, counted once per task graph. \lctes~ran again for this comparison, and its total time differs by $5\%$ from the one in the single-PE comparison. The time ratios count the time of \lctes~in every configuration, since \mastag~needs it to schedule under any limit. The limit of 600 seconds on the schedulers of \mastag~is reached in 202 calls, on montage (171) and cycles (31): this shortens the running time of \mastag~and may leave \mastag~with a longer makespan on these instances. On the 2 cycles graphs where \lctes~returns no schedule, the schedulers of \mastag~return none either, under every limit, and we count no time for them.

Table~\ref{tab:vs-mastag-norefine} gives the results of \mastag~without the refinement. \mastag~still schedules all instances under the M limits, and its parallel schedulers are about $4$ times faster than with the refinement, on geometric mean. Its makespans are within $0.2\%$ of those with the refinement, on geometric mean, except on montage, where they are $11\%$ shorter, mainly since only 22 calls stop at 600 seconds instead of 171. The portfolio is $2.9$ to $4.1$ times faster than \mastag, and its makespan is within $0.2\%$ to $4.6\%$ of that of \mastag, on median.

\begin{table}[ht]
\centering
\small
\caption{Portfolio vs \mastag~(refinement on) on 4 and 8 PEs: share of instances scheduled within the limit, makespan ratio (ours divided by \mastag, geometric mean over the instances both schedule), and time ratio (\mastag~divided by ours, geometric mean; in parentheses, without \lctes). The time ratio is over all instances with an \lctes~schedule.
}
\label{tab:vs-mastag-multi}
\begin{tabular}{l|cc|cc|cc}
\hline
\multirow{2}{*}{Limit} & \multicolumn{2}{c|}{Scheduled, \% (ours / \mastag)} & \multicolumn{2}{c|}{Makespan ratio} & \multicolumn{2}{c}{Time ratio} \\
 & 4 PEs & 8 PEs & 4 PEs & 8 PEs & 4 PEs & 8 PEs \\
\hline
M & 63.4 / 100 & 63.1 / 100 & 1.074 & 1.100 & 29.8 (10.3) & 22.5 (7.9) \\
M$+25\%$ & 99.7 / 100 & 99.7 / 100 & 1.044 & 1.087 & 21.8 (5.6) & 17.8 (4.6) \\
M$+50\%$ & 100 / 100 & 100 / 100 & 1.030 & 1.066 & 23.1 (5.6) & 18.9 (4.7) \\
M$+75\%$ & 100 / 100 & 100 / 100 & 1.017 & 1.033 & 25.6 (6.3) & 21.4 (5.4) \\
G & 100 / 93.5 & 100 / 93.5 & 1.123 & 1.170 & 25.2 (5.2) & 19.9 (4.2) \\
G$+0.1\%$ & 100 / 93.5 & 100 / 93.5 & 1.101 & 1.136 & 23.0 (4.8) & 18.2 (3.9) \\
G$+1\%$ & 100 / 94.7 & 100 / 94.7 & 1.071 & 1.106 & 22.9 (5.1) & 17.9 (4.1) \\
\hline
\end{tabular}
\end{table}

\begin{table}[ht]
\centering
\small
\caption{Portfolio vs \mastag~(refinement off) on 4 and 8 PEs: makespan and time ratios as in Table~\ref{tab:vs-mastag-multi}}
\label{tab:vs-mastag-norefine}
\begin{tabular}{l|cc|cc}
\hline
\multirow{2}{*}{Limit} & \multicolumn{2}{c|}{Makespan ratio} & \multicolumn{2}{c}{Time ratio} \\
 & 4 PEs & 8 PEs & 4 PEs & 8 PEs \\
\hline
M & 1.079 & 1.111 & 13.2 (1.6) & 10.1 (1.3) \\
M$+25\%$ & 1.046 & 1.090 & 12.3 (1.5) & 10.1 (1.3) \\
M$+50\%$ & 1.033 & 1.065 & 13.1 (1.6) & 11.3 (1.4) \\
M$+75\%$ & 1.025 & 1.048 & 13.9 (1.8) & 11.7 (1.5) \\
G & 1.125 & 1.176 & 12.1 (1.2) & 9.7 (1.0) \\
G$+0.1\%$ & 1.104 & 1.143 & 11.3 (1.1) & 9.1 (1.0) \\
G$+1\%$ & 1.075 & 1.115 & 11.3 (1.2) & 8.9 (1.0) \\
\hline
\end{tabular}
\end{table}

\begin{table}[ht]
\centering
\small
\caption{Portfolio vs \mastag~per workflow, over the 7 limits: share of instances scheduled within the limit (in \%, ours / \mastag, over 4 and 8 PEs), makespan ratio (ours divided by \mastag, geometric mean over the instances both schedule), and time ratio (\mastag~divided by ours, geometric mean; in parentheses, without \lctes). Our failures are under the M limits and those of \mastag~under the G limits, so the makespan ratio of each workflow is over a different mix of limits.}
\label{tab:vs-mastag-workflow}
\setlength{\tabcolsep}{3pt}
\begin{tabular}{l|r|c|cc|cc}
\hline
\multirow{2}{*}{Workflow} & \multirow{2}{*}{Graphs} & Scheduled, \% & \multicolumn{2}{c|}{Makespan ratio} & \multicolumn{2}{c}{Time ratio} \\
 & & (ours / \mastag) & 4 PEs & 8 PEs & 4 PEs & 8 PEs \\
\hline
1000genome & 22 & 86.4 / 100 & 1.139 & 1.264 & 441 (5.6) & 265 (3.3) \\
blast & 15 & 100 / 100 & 1.004 & 1.019 & 4.9 (4.4) & 4.5 (4.1) \\
bwa & 15 & 100 / 100 & 1.000 & 1.001 & 7.5 (6.5) & 6.9 (6.0) \\
cycles & 24 & 100 / 61.3 & 0.997 & 1.048 & 1,773 (5.1) & 1,439 (4.0) \\
epigenomics & 26 & 95.6 / 100 & 1.012 & 1.027 & 5.3 (4.1) & 4.2 (3.2) \\
montage & 17 & 97.5 / 100 & 1.257 & 1.457 & 113 (98.4) & 75.2 (66.1) \\
nextflow & 15 & 92.9 / 100 & 1.076 & 1.081 & 392 (8.8) & 347 (9.9) \\
synth-genome & 42 & 92.5 / 100 & 1.099 & 1.157 & 1.8 (1.5) & 1.4 (1.1) \\
synth-ligo & 42 & 93.2 / 100 & 1.055 & 1.072 & 66.1 (7.6) & 53.0 (6.3) \\
synth-montage & 42 & 100 / 100 & 1.001 & 1.001 & 6.4 (4.5) & 5.3 (3.6) \\
RASflow & 7 & 100 / 100 & 1.046 & 1.057 & 12.2 (11.0) & 10.1 (9.2) \\
rna-seq-star-deseq2 & 19 & 87.2 / 100 & 1.065 & 1.098 & 37.4 (7.6) & 33.6 (7.1) \\
seismology & 11 & 100 / 100 & 1.000 & 1.000 & 8.2 (7.8) & 7.5 (7.1) \\
soykb & 10 & 97.1 / 100 & 1.071 & 1.099 & 183 (28.3) & 154 (24.7) \\
srasearch & 25 & 85.7 / 100 & 1.203 & 1.239 & 5.1 (4.1) & 3.9 (3.1) \\
streamflow & 1 & 100 / 100 & 1.000 & 1.000 & 31.1 (2.1) & 26.7 (1.8) \\
varlociraptor & 8 & 100 / 100 & 1.000 & 1.000 & 57.0 (14.7) & 47.6 (12.9) \\
\hline
All & 341 & 94.7 / 97.4 & 1.064 & 1.098 & 24.4 (6.0) & 19.4 (4.8) \\
\hline
\end{tabular}
\end{table}

\paragraph{Online scheduling.} Our algorithm uses the whole task graph: it places each task in the earliest idle window that fits, possibly before tasks that come earlier in the order. When tasks become known only once they are ready, e.g., in a runtime system, this knowledge is not available. To see how our algorithm does without this knowledge, we also run it online: whenever a task ends, each idle PE starts the first ready task in the order that fits in memory, and a PE never waits for a later task. \mastag~also starts tasks whenever a task ends, but it uses its sequential schedule to keep enough memory for the remaining tasks, and its orders need the whole graph, so it cannot run online. Our online run still respects the memory limit, since a task starts only if it fits. We use five task sorting methods whose keys are known once a task is ready, \wspp, \ws, DepthMaxMem, DepthMax, and DepthMin, one run each with the PE rule Load, on the same instances and limits, see Table~\ref{tab:online}. Under M, which leaves no slack above the sequential peak of \lctes, a single online run schedules only $25\%$ to $38\%$ of the instances, but a modest slack is enough: it schedules $84\%$ to $98\%$ under M$+25\%$ to M$+75\%$. Its makespan is mostly shorter than the one of the static run with the same order and PE rule, since it never leaves a PE idle while a ready task fits, whereas the static run places tasks in priority order and may leave gaps. On the instances it schedules, the makespan of the online run is $9\%$ shorter to $2\%$ longer than the one of the static portfolio, and $0\%$ to $7\%$ longer than the one of \mastag. These ratios favour the online run, which schedules mostly the easier instances. Under the G limits, it schedules $50\%$ to $92\%$ of the instances, fewer than the static runs, since it starts any ready task that fits and cannot keep memory free for a later one. Note that the tie-break $u(\cdot)$ of \wspp~uses the data waiting for the children of a task, i.e., information one step beyond the ready tasks, so \wspp~is not strictly online. The other four methods are strictly online, and \ws~gives almost the same results.

\begin{table}[ht]
\centering
\small
\caption{Online runs, one task sorting method each (PE rule Load), on 4 and 8 PEs: share of instances scheduled (range over the limits of each family and both PE counts), and makespan ratio (geometric mean over the instances both schedule; range over the 14 configurations) to the static run of the same method and PE rule, to the static portfolio, and to \mastag}
\label{tab:online}
%\resizebox{\columnwidth}{!}{
\begin{tabular}{l|ccc|ccc}
\hline
\multirow{2}{*}{Method} & \multicolumn{3}{c|}{Feasible (\%)} & \multicolumn{3}{c}{Makespan ratio to} \\
 & M & M$+25$--$75\%$ & G family & static, same method & portfolio & \mastag \\
\hline
\wspp & 33--34 & 91--98 & 69--92 & 0.85--0.95 & 0.95--1.01 & 1.00--1.05 \\
\ws & 34 & 89--97 & 69--92 & 0.85--0.94 & 0.95--1.01 & 1.01--1.05 \\
DepthMaxMem & 38 & 85--91 & 57--80 & 0.77--0.93 & 0.95--1.01 & 1.03--1.07 \\
DepthMax & 25 & 84--92 & 53--80 & 0.74--0.94 & 0.94--1.02 & 1.01--1.05 \\
DepthMin & 26 & 84--92 & 50--77 & 0.83--0.99 & 0.91--0.99 & 1.00--1.03 \\
\hline
\end{tabular}
\end{table}

\subsection{Deep Learning}\label{sec:experiments-deep-learning}

\paragraph{Setup.} We integrated Algorithm~\ref{alg:mem-aware-graph} in the backend of the MindSpore 2.6.0 \cite{chen2021deep} open-source deep learning framework. 
We ran training tests on a Kunpeng 920 CPU 2.6 GHz server featuring 8 Ascend 910B1 accelerators \cite{liao2021ascend},
each featuring 64 GB high bandwidth memory.

\paragraph{Workload I.} We experimented with the training of \textit{Llama3.1-8b} \cite{llama3.1} (about 9,000 tasks per graph) and \textit{TeleChat2-7b} \cite{wang2025technical} (about 13,000 tasks per graph). We considered a variety of configurations capturing combinations of data parallelism (DP), model parallelism (MP), and MindSpore compilation optimization levels: O0 is the default MindSpore compilation and O1 is the advanced mode, where, for example, task kernels are fused to foster efficiency. We assigned costs based on timing information obtained from advanced profiling and partitioned tasks into a computation and a communication type (each corresponding to a single PE). 

We use the three heuristics TIC \cite{hashemi2019tictac}, TAC \cite{hashemi2019tictac}, and S/C \cite{li2023s}, recall (xii) from Section~\ref{sec:task-sorting}, as baselines. TIC and TAC are seminal scheduling ideas in the deep learning systems literature, which pay particular attention to communication operators for parallel training, while S/C implements memory-aware tie-breaking for depth-first scheduling in the context of query processing. Note that other scheduling approaches (for deep learning or other industrial settings discussed in related work) do not only work at the algorithmic level, but also perform other optimizations like graph rewriting. On the other hand, our method is model-agnostic, orthogonal to any graph rewriting optimization, and does not affect model convergence (loss function). To enable a fair comparison in the memory-aware context we consider, we implemented the three baselines inside our approach as task sorting methods. 
In other words, all methods in Tables~\ref{tab:llama-speedup}--\ref{tab:telechat-memory} use the same placement and memory checks of Algorithm~\ref{alg:mem-aware-graph}, with the same memory limit; only the order in which ready tasks are considered differs. TIC gives priority to the transfers that unblock computation the earliest, from the structure of the graph, while TAC uses the estimated times to give priority to the transfers that shorten the remaining computation the most. Both aim to overlap communication with computation as much as possible \cite{hashemi2019tictac}. S/C follows a depth-first order, where ties are broken by memory \cite{li2023s}.

For our algorithm portfolio, we run the task sorting methods of an earlier version of our implementation with the Load PE rule and keep the schedule with the shortest estimated makespan: these are the methods of Section~\ref{sec:task-sorting} without \wspp, and with the variants of \ws~that we omit there. The experiments in this subsection precede \wspp~and the portfolio of four runs of Section~\ref{sec:experiments-scientific-workflows}.

The schedule is computed on the host CPU during graph compilation. Then, all tasks of the training step run on the Ascend accelerators: each accelerator executes the tasks of its own graph, with one computation and one communication stream.

In Tables~\ref{tab:llama-speedup} and \ref{tab:telechat-speedup}, we present the  speedups obtained when we set $\wmax$ equal to the device memory. \textit{Default} corresponds to the manually fine-tuned schedule in MindSpore default mode. Speedup is the median training step time of the compared method divided by ours, over $40$ training steps. In parentheses, we show the expected speedup based on the profiling-cost estimated makespan for each schedule: due to system delays, e.g., in synchronizations, communication, or inefficient computation and communication kernel overlapping, the theoretical speedup is usually an upper bound on the real one. On the DP8 MP1 configurations, the real speedup may exceed the estimate (up to $1.08$ against $1.00$, on TeleChat2). The results show that we obtain up to $1.10$x speedups for training step time.

\begin{table}[ht]
\centering
\small
\caption{Algorithm~\ref{alg:mem-aware-graph} speedups for Llama3.1}
\label{tab:llama-speedup}
\begin{tabular}{l|cccccc}
\hline
\multirow{2}{*}{Configuration} & 
\multicolumn{4}{c}{Real (Theoretical) Speedup} \\
\cline{2-5}
 & Over Default & Over TIC \cite{hashemi2019tictac} & Over TAC \cite{hashemi2019tictac} & Over S/C \cite{li2023s} \\
\hline
DP8 MP1 O0 & 1.02 (1.01) & 1.01 (1.00) & 1.00 (0.99) & 1.05 (1.10) \\
DP8 MP1 O1 & 1.01 (1.00) & 1.00 (0.99) & 1.00 (0.99) & 1.03 (1.06) \\
DP4 MP2 O0 & 1.09 (1.13) & 1.08 (1.12) & 1.07 (1.12) & 1.10 (1.18) \\
DP4 MP2 O1 & 1.09 (1.13) & 1.08 (1.12) & 1.08 (1.12) & 1.08 (1.14) \\
DP2 MP4 O0 & 1.07 (1.11) & 1.05 (1.11) & 1.07 (1.11) & 1.10 (1.18) \\
DP2 MP4 O1 & 1.10 (1.11) & 1.09 (1.11) & 1.09 (1.11) & 1.10 (1.15) \\
DP1 MP8 O0 & 1.07 (1.11) & 1.06 (1.11) & 1.06 (1.11) & 1.10 (1.18) \\
DP1 MP8 O1 & 1.10 (1.11) & 1.08 (1.11) & 1.08 (1.11) & 1.10 (1.15) \\
\hline
\end{tabular}
\end{table}

\begin{table}[ht]
\centering
\caption{Algorithm~\ref{alg:mem-aware-graph} speedups for TeleChat2}
\label{tab:telechat-speedup}
\begin{tabular}{l|cccccc}
\hline
\multirow{2}{*}{Configuration} & 
\multicolumn{4}{c}{Real (Theoretical) Speedup} \\
\cline{2-5}
 & Over Default & Over TIC \cite{hashemi2019tictac} & Over TAC \cite{hashemi2019tictac} & Over S/C \cite{li2023s} \\
\hline
DP8 MP1 O0 & 1.08 (1.00) & 1.04 (1.00) & 1.05 (1.00) & 1.08 (1.06) \\
DP8 MP1 O1 & 1.05 (1.02) & 1.01 (1.01) & 1.01 (1.01) & 1.07 (1.08) \\
DP4 MP2 O0 & 1.05 (1.10) & 1.03 (1.09) & 1.03 (1.08) & 1.05 (1.14)\\
DP4 MP2 O1 & 1.05 (1.11) & 1.04 (1.10) & 1.04 (1.10) & 1.07 (1.15) \\
DP2 MP4 O0 & 1.03 (1.10) & 1.04 (1.09) & 1.04 (1.09) & 1.06 (1.15) \\
DP2 MP4 O1 & 1.03 (1.11) & 1.04 (1.10) & 1.04 (1.11) & 1.07 (1.19) \\
DP1 MP8 O0 & 1.06 (1.10) & 1.06 (1.09) & 1.07 (1.09) & 1.04 (1.06)\\
DP1 MP8 O1 & 1.07 (1.11) & 1.06 (1.10) & 1.05 (1.10) & 1.05 (1.06) \\
\hline
\end{tabular}
\end{table}

Besides obtaining useful speedups, our algorithm needs at most $3.62\%$ more memory than Default (Llama3.1 DP2 MP4 O1 and DP1 MP8 O1, with a $1.10$x speedup), and at most $0.73\%$ more in the other configurations, see Tables~\ref{tab:llama-memory} and \ref{tab:telechat-memory}. In the best case (TeleChat2 DP1 MP8 O0), our algorithm saves $18.05\%$ in memory, while also speeding up at least $1.04$x. Memory is the peak measured on the device.

\begin{table}[ht]
\centering
%\small
\caption{Algorithm~\ref{alg:mem-aware-graph} relative memory for Llama3.1}
\label{tab:llama-memory}
%\resizebox{\columnwidth}{!}{
\begin{tabular}{l|ccccc}
\hline
\multirow{2}{*}{Configuration} & 
\multicolumn{4}{c}{Relative Memory Difference} \\
\cline{2-5}
 & w.r.t. Default & w.r.t. TIC \cite{hashemi2019tictac} & w.r.t. TAC \cite{hashemi2019tictac} & w.r.t. S/C \cite{li2023s} \\
\hline
DP8 MP1 O0 & +0.00\% & +0.00\% & +0.00\% & +0.00\% \\
DP8 MP1 O1 & +0.00\% & +0.00\% & +0.00\% & -0.56\%  \\
DP4 MP2 O0 & +0.00\% & +0.00\% & +0.00\% & -0.24\%  \\
DP4 MP2 O1 & +0.17\% & +0.17\% & +0.00\% & +0.17\% \\
DP2 MP4 O0 & -8.27\% & -8.27\% & -8.27\% & -1.99\% \\
DP2 MP4 O1 & +3.62\% & +0.00\% & -0.10\% & +0.00\%  \\
DP1 MP8 O0 & -11.47\% & -11.47\% & -11.47\% & -5.41\%  \\
DP1 MP8 O1 & +3.62\% & +0.00\% & -0.22\% & +0.00\%  \\
\hline
\end{tabular}
%}
\end{table}

\begin{table}[ht]
\centering
%\small
\caption{Algorithm~\ref{alg:mem-aware-graph} relative memory for TeleChat2}
\label{tab:telechat-memory}
%\resizebox{\columnwidth}{!}{
\begin{tabular}{l|ccccc}
\hline
\multirow{2}{*}{Configuration} & 
\multicolumn{4}{c}{Relative Memory Difference} \\
\cline{2-5}
 & w.r.t. Default & w.r.t. TIC \cite{hashemi2019tictac} & w.r.t. TAC \cite{hashemi2019tictac} & w.r.t. S/C \cite{li2023s} \\
\hline
DP8 MP1 O0 & +0.00\% & +0.00\% & -4.35\% & +0.00\% \\
DP8 MP1 O1 & +0.00\% & +0.00\% & +0.00\% & +0.00\% \\
DP4 MP2 O0 & +0.00\% & +0.00\% & +0.00\% & -3.18\% \\
DP4 MP2 O1 & +0.73\% & +0.73\% & +0.73\% & -2.71\% \\
DP2 MP4 O0 & +0.00\% & -10.00\% & -9.67\% & +0.00\% \\
DP2 MP4 O1 & +0.42\% & -2.24\% & +0.42\% & -0.50\% \\
DP1 MP8 O0 & -18.05\% & -18.05\% & -18.05\% & -18.05\% \\
DP1 MP8 O1 & +0.00\% & +0.00\% & +0.00\% & -1.66\% \\
\hline
\end{tabular}
%}
\end{table}

So far, among the schedules of all task sorting methods, we kept the one with the shortest estimated makespan. Instead, we may keep the one with the lowest peak memory, breaking ties (within $1\%$ of memory) by makespan. For TeleChat2, the selected schedules are the same. For Llama3.1, only the two configurations with $+3.62\%$ memory over Default change, see Table~\ref{tab:llama-memory-select}: the extra memory disappears, and our schedule needs $3.49\%$ to $3.71\%$ less memory than TIC, TAC, and S/C, but the speedup over Default drops from $1.10$x to $1.02$x and $0.99$x, respectively. In other words, the portfolio offers a choice between makespan and memory, and the user may pick either.

\begin{table}[ht]
\centering
\caption{Llama3.1, selecting the schedule with the lowest peak memory}
\label{tab:llama-memory-select}
%\resizebox{\columnwidth}{!}{
\begin{tabular}{l|cccc}
\hline
Configuration & Over Default & Over TIC \cite{hashemi2019tictac} & Over TAC \cite{hashemi2019tictac} & Over S/C \cite{li2023s} \\
\hline
\multicolumn{5}{l}{Real (theoretical) speedup} \\
DP2 MP4 O1 & 1.02 (1.02) & 1.01 (1.02) & 1.01 (1.02) & 1.02 (1.06) \\
DP1 MP8 O1 & 0.99 (0.90) & 0.98 (0.89) & 0.98 (0.89) & 0.99 (0.93) \\
\hline
\multicolumn{5}{l}{Relative memory difference} \\
DP2 MP4 O1 & +0.00\% & -3.49\% & -3.59\% & -3.49\% \\
DP1 MP8 O1 & +0.00\% & -3.49\% & -3.71\% & -3.49\% \\
\hline
\end{tabular}
%}
\end{table}

\paragraph{Workload II.} We experimented with the data parallel training of \textit{Mixtral-8x7b} \cite{jiang2024mixtral} (DP8 MP1 O0) with 8 layers.
Each accelerator receives the same graph (roughly 12,000 tasks), where we assign task costs based on a mathematical model of operator intensity. We assume two PEs, a computation and a communication one, and assign tasks to them accordingly.
In Table~\ref{tab:mixtral}, we showcase how our algorithm reacts to memory pressure.
Peak memory is given in megabytes and training step time in milliseconds, where we show the median of 100 steps.
By running Algorithm~\ref{alg:mem-aware-graph} with a decreasing memory limit, we are able to reduce peak memory and step time at the same time, compared to the default manually-tuned approach in MindSpore.
A simple graph-based memory lower bound is 38131 megabytes, and indeed, Algorithm~\ref{alg:mem-aware-graph} returns None when $\wmax = 37888 < 38131$. 

\begin{table}[ht]
    \centering
    \caption{Experimental result for Mixtral-8x7b training}
    \label{tab:mixtral}
    %\resizebox{\columnwidth}{!}{
    \begin{tabular}{l|l|l|l}
        \hline
         Method & $\wmax$ & Peak Memory  & Step Time\\\hline
         Default &-- & 46364 & 2720
         \\
         Algorithm~\ref{alg:mem-aware-graph} & 45056 & 44230  & 2647 \\
         Algorithm~\ref{alg:mem-aware-graph} & 44032 & 38131 & 2631 \\
         Algorithm~\ref{alg:mem-aware-graph} & 37888 & None & None \\
         \hline
    \end{tabular}
    %}
\end{table}

Overall, inside MindSpore, our \texttt{C++} implementation gives training speedups of up to $1.10$x over Default with at most $3.62\%$ more memory, and under memory pressure it reduces both peak memory and step time. Next, we compare with \mastag~on deep learning graphs, outside MindSpore, since \mastag~exists only in \texttt{Python}.

\paragraph{Scaling: comparison with \mastag.} We run the \texttt{Python} implementation of Section~\ref{sec:experiments-scientific-workflows} on 7 training graphs of large language models from MindSpore, converted to the DAX format, with 2,354 to 115,528 tasks, see Table~\ref{tab:dl-scaling}. Three graphs come from DeepSeek-V3 \cite{deepseekai2024deepseekv3}, a mixture-of-experts model: two reduced to 6 and 8 layers, with DP4 MP2 and DP2 MP4, respectively, and one in a pipeline-parallel configuration (PP). Three come from Llama3.1, at the compilation levels O1 and O0, and with 4 interleaved micro-batches (MB4). The last one comes from TeleChat2. The Llama3.1 and TeleChat2 graphs were captured separately from those of Workloads I and II, for parallel configurations that the converted graphs do not record. So, their numbers are not comparable with those of Workloads I and II. We compare \wspp~on a single PE with \lctes, with the same settings as in Section~\ref{sec:experiments-scientific-workflows}. As there, any task may run on any PE with the same cost, i.e., we ignore the computation and communication types of the tasks, since \mastag~assumes identical PEs. \lctes~reports a time-out on 3 of the 5 graphs it schedules, runs for more than 600 seconds on all 5 graphs it schedules, and returns no schedule within a day on the 2 largest, while \wspp~takes at most 33 seconds, with times as in Section~\ref{sec:experiments-scientific-workflows}. Where \lctes~returns a schedule, \wspp~has the same peak memory within $0.01\%$, except on Llama3.1 O0, where it is $18\%$ higher, as it postpones the reductions of the overflow check and of the gradient norm. On this graph, five of our 17 methods are within $0.01\%$ of \lctes.
In Table~\ref{tab:dl-multi}, we compare the portfolio with \mastag~on 4 and 8 PEs, with the limits of Section~\ref{sec:experiments-scientific-workflows}: our portfolio schedules none of the 5 graphs with an \lctes~schedule under M, 4 of them under M$+25\%$ to M$+75\%$, and all 7 graphs under the G limits, where \mastag~schedules 5. Here, we run \mastag~without its refinement, since with it the schedulers (ii) and (iii) stop at 600 seconds on every graph. On the graphs both tools schedule, the makespans of the two tools are within $1\%$ of each other on geometric mean. Under the limits other than M, the portfolio is $217$ to $332$ times faster than \mastag~on geometric mean, as it does not need \lctes, and $4.2$ to $7.7$ times faster than the parallel schedulers of \mastag~alone. No run of the portfolio stops at our time limit of 600 seconds.

\begin{table}[ht]
\centering
\small
\caption{Deep learning graphs on a single PE: running time and peak memory of \wspp~against \lctes~(peak ratio = \wspp~/ \lctes); $^*$: \lctes~reports a time-out and returns its best schedule so far; --: no schedule within a day}
\label{tab:dl-scaling}
\begin{tabular}{l|r|r|r|c}
\hline
Graph & Tasks & \wspp~(s) & \lctes~(s) & Peak ratio \\
\hline
DeepSeek-V3 DP4 MP2 L6 & 2,354 & 0.33 & 779$^*$ & 1.0001 \\
DeepSeek-V3 DP2 MP4 L8 & 3,076 & 0.47 & 1,044$^*$ & 1.0001 \\
Llama3.1 O1 & 6,092 & 1.24 & 1,941 & 1.0000 \\
Llama3.1 O0 & 8,614 & 2.26 & 7,886$^*$ & 1.1829 \\
TeleChat2 & 13,291 & 2.76 & 4,102 & 1.0000 \\
Llama3.1 MB4 & 18,479 & 4.01 & -- & -- \\
DeepSeek-V3 PP & 115,528 & 32.83 & -- & -- \\
\hline
\end{tabular}
\end{table}

\begin{table}[ht]
\centering
\caption{Deep learning graphs on 4 and 8 PEs: our portfolio of four pairs against \mastag~(refinement off); makespan ratio = \mastag~/ ours~(geometric mean over the graphs both schedule); time ratio = \mastag~/ ours, with \lctes~(without it in parentheses), over the 5 graphs with an \lctes~schedule; --: our portfolio schedules none.}
\label{tab:dl-multi}
\begin{tabular}{l|c|c|cc|cc}
\hline
\multirow{2}{*}{Limit} & \multirow{2}{*}{Graphs} & Scheduled & \multicolumn{2}{c|}{Makespan ratio} & \multicolumn{2}{c}{Time ratio} \\
 & & Ours / \mastag & 4 PEs & 8 PEs & 4 PEs & 8 PEs \\
\hline
M & 5 & 0 / 5 & -- & -- & -- & -- \\
M$+25\%$ & 5 & 4 / 5 & 1.00 & 1.00 & 305 (5.9) & 227 (4.3) \\
M$+50\%$ & 5 & 4 / 5 & 1.00 & 1.00 & 314 (5.9) & 232 (4.5) \\
M$+75\%$ & 5 & 4 / 5 & 1.00 & 1.00 & 296 (5.7) & 217 (4.2) \\
G & 7 & 7 / 5 & 1.00 & 1.01 & 323 (7.7) & 238 (6.3) \\
G$+0.1\%$ & 7 & 7 / 5 & 1.00 & 1.01 & 317 (7.0) & 248 (5.8) \\
G$+1\%$ & 7 & 7 / 5 & 1.00 & 1.00 & 332 (6.6) & 252 (5.1) \\
\hline
\end{tabular}
\end{table}

\paragraph{Launch order and synchronizations.}
As a side contribution, we present a way to force the target hardware, e.g., an AI accelerator, to respect the  precedence constraints imposed by a schedule generated by Algorithm~\ref{alg:mem-aware-graph}, regardless of fluctuations in real task costs during runtime. That is, if in the generated schedule a task starts after the finish time of another task, we wish for this to happen in practice as well. This way, the memory consumption in $\peak$ is respected at any given time and we avoid out-of-memory errors in practice. Assume an interface enables us to pass to the device a task launch order, which also includes stream synchronizations in the form of wait events, like in our setup with MindSpore and Ascend. 
Tasks are launched in ascending order of start time in the schedule, and, ideally, we wish for the minimum number of synchronizations to avoid incurring overhead. Current systems allow us to manually define the number of execution streams: even if in reality there are only a few PEs available, we may define multiple execution streams. Initially, we let a PE correspond to a single execution stream.

\begin{figure}
    \centering    \includegraphics[width=0.5\linewidth]{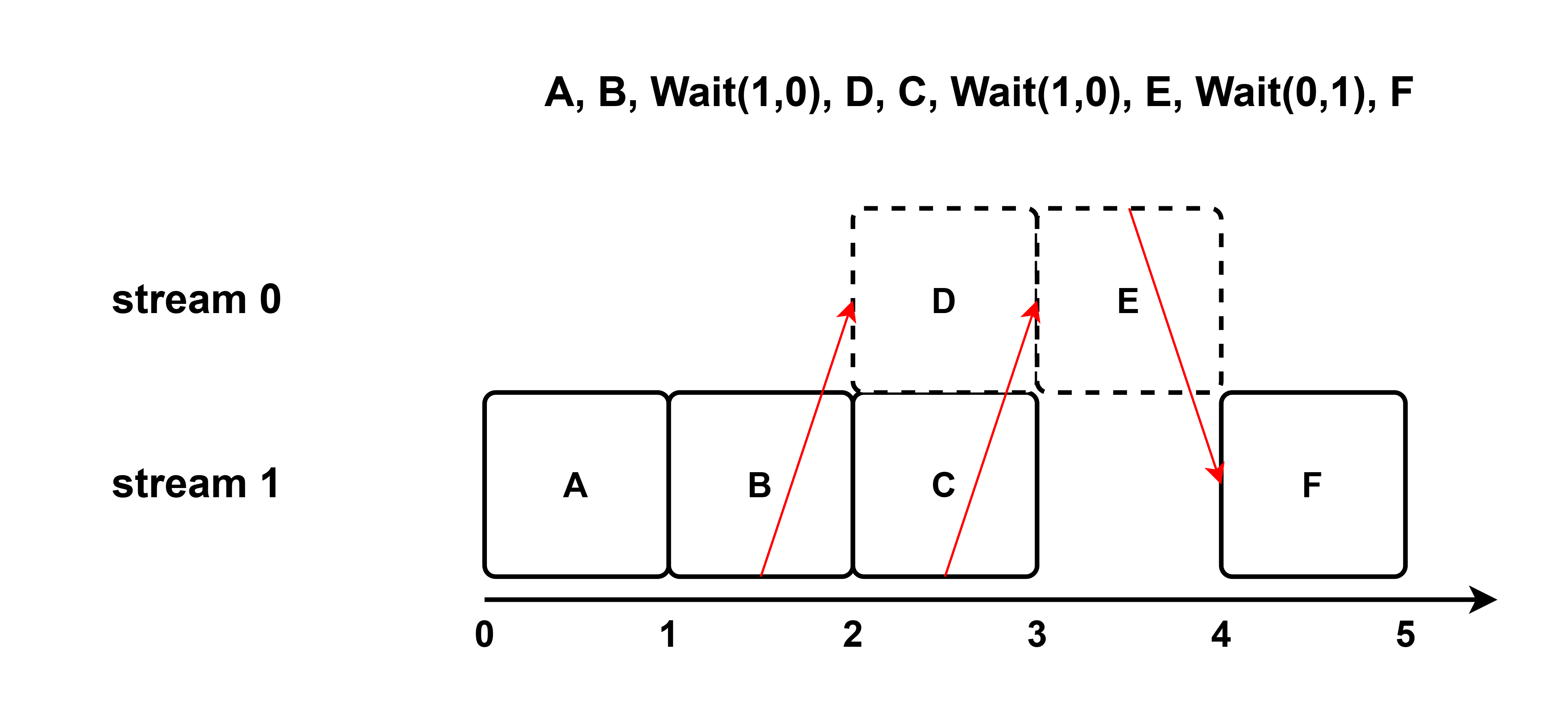}
    \caption{Transforming a schedule into a launch order: Tasks $A$ and $B$ are launched. Before $D$ is executed it must wait for $B$ to finish, due to the $\mathit{Wait}(1,0)$ synchronization. Then, $C$ is launched and can run parallel to $D$. Before $E$ executes it must wait for the latest stream $1$ task, namely $C$. Finally, $F$ will wait to start execution after $E$ due to the $\mathit{Wait}(0,1)$ synchronization.}
    \label{fig:naive-events}
\end{figure}

The first idea is to add a synchronization between a task and all tasks in its immediate right neighborhood in terms of time and assigned to a different stream. For $n_1 \in N$,  let $R(n_1) \defeq \{ n_2 \in N\;|\; \sigma(n_2) \ge \sigma(n_1) + c(n_1, p(n_1))\}$ denote the set of tasks starting after $n_1$ finishes and $r(n_1) \defeq \min_{n_2 \in R(n_1)} \sigma(n_2) + c(n_2, p(n_2))$ denote the minimum finish time among them. We define the \textit{sync neighborhood} of $n_1$ to be $S(n_1) \defeq \{n_2 \in R(n_1)\;|\; \sigma(n_2) < r(n_1) \,\land\, p(n_1) \neq p(n_2)\}$. Intuitively, we wish to enforce all tasks in $S(n_1)$ to start execution only after $n_1$ finishes, as dictated by the schedule. Consider $n_2 \in R(n_1)\setminus S(n_1)$.
If $p(n_1) = p(n_2)$, then synchronization is not necessary, since each stream executes tasks launched on it sequentially. If $p(n_1) \neq p(n_2)$, then by the above definitions, there exists a task $n_3 \in N$, where $n_3 \neq n_1$, $n_3 \neq n_2$, and $\sigma(n_3) + c(n_3, p(n_3)) = r(n_1)$, such that it holds $r(n_1) \le \sigma(n_2)$. To enforce $n_2$ to wait for $n_1$, there is no need for an extra synchronization event, as based on the sync neighborhood argument, $n_3$ will wait for $n_1$ and $n_2$ will wait for $n_3$. In Figure~\ref{fig:naive-events}, we show a transformation from a schedule to a launch order for the example in Figure~\ref{fig:example}; $\mathit{Wait(X,Y)}$ denotes that stream $Y$ must wait for stream $X$ to finish its latest launched task execution.

\begin{figure*}[ht]
    \centering
    \includegraphics[width=0.85\linewidth]{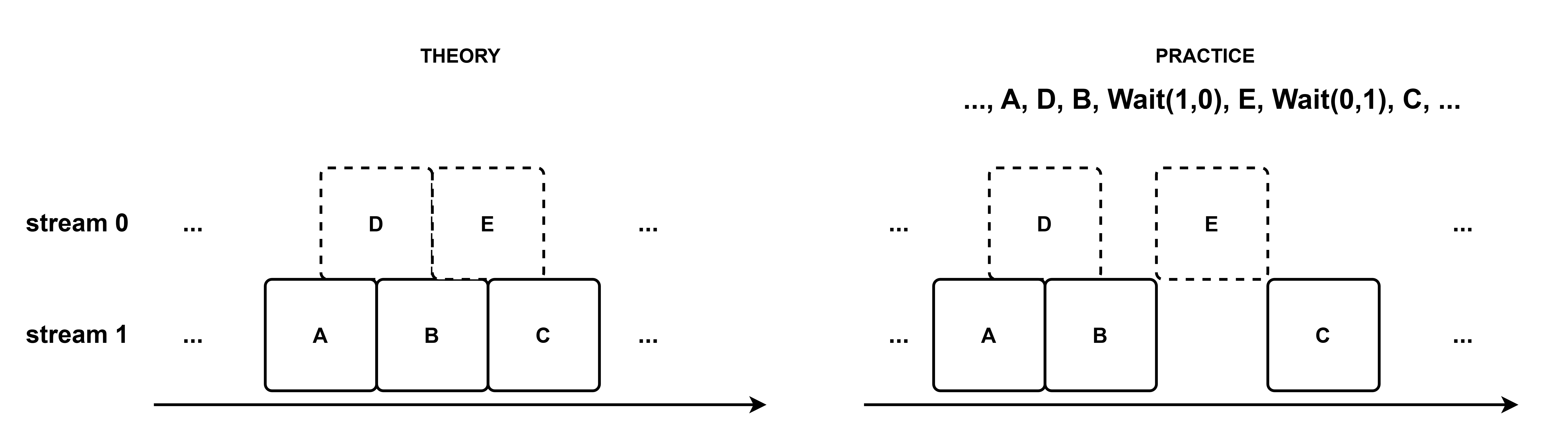}
    \caption{Slowdown due to first PE synchronization approach within a task launch order}
    \label{fig:naive-events-limitation}
\end{figure*}

In Figure~\ref{fig:naive-events}, we are able to respect the theoretical schedule with the given launch order. Note that another launch order prioritizing $C$ instead of $D$ would serialize the execution of $C$ and $D$, due to the $\mathit{Wait}(1,0)$ event. In general, following the sync neighborhood rule may lead to slowdowns in practice, regardless of tie-breaking for tasks scheduled at the same time, since wait events are tied to streams and not to specific tasks. In the partial schedule given in Figure~\ref{fig:naive-events-limitation}, the constraints imposed by the schedule essentially dictate that $E$ must wait for $A$ and $C$ must wait for $D$; otherwise, memory safety is not guaranteed. However, since $B$ is launched before $E$, a $\mathit{Wait}(1,0)$ event forces $E$ to wait for $B$ as well. Similarly, $C$ waits for $E$, because $\mathit{Wait}(0,1)$ applies to task $E$ in stream $0$.

\begin{figure*}[ht]
    \centering
    \includegraphics[width=0.75\linewidth]{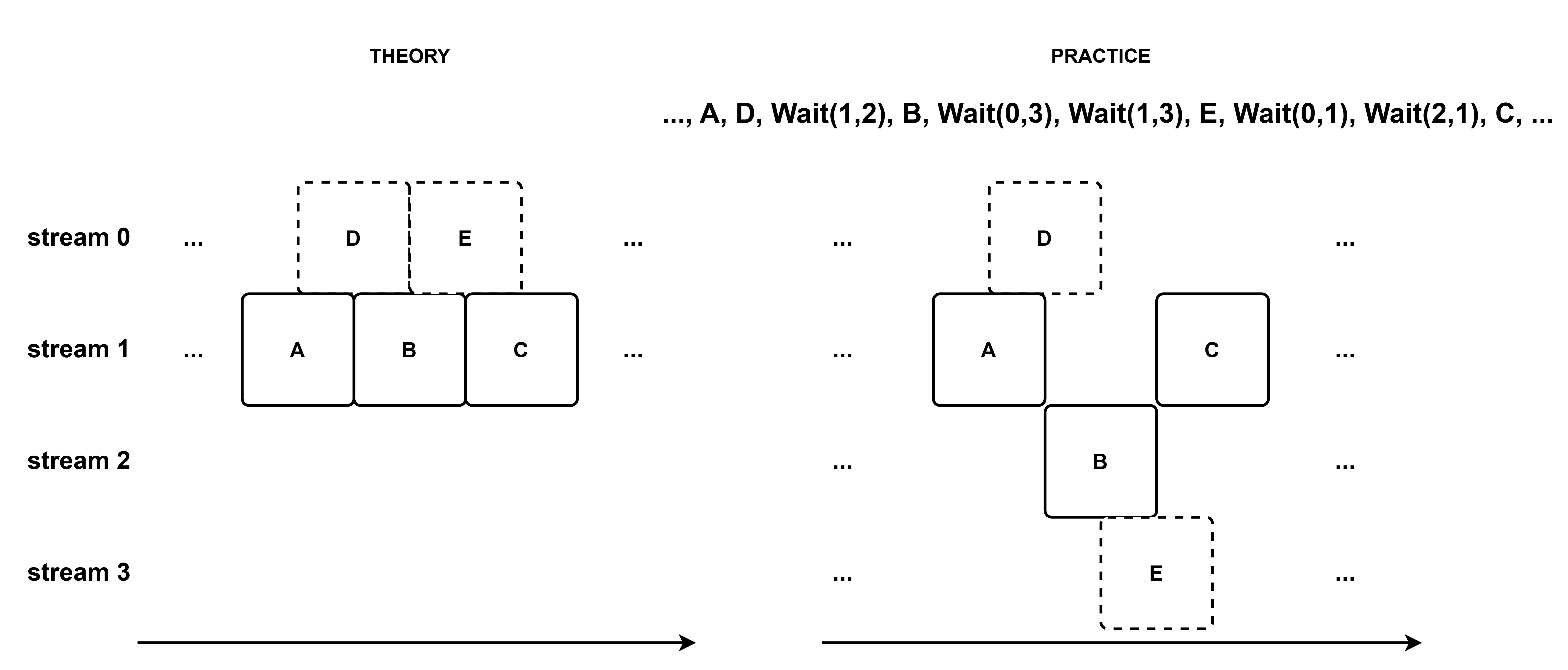}
    \caption{Harnessing slowdowns via PE space lifting and corresponding wait events in task launch orders}
    \label{fig:lift-events}
\end{figure*}

We devise a novel approach to tackle the above limitation. To represent task-tied synchronizations in the same interface, we lift the space of available streams. For each ordered pair of original streams $(P_i,P_j) \in P\times P$, where $P_i \neq P_j$, we define a new stream $P_{i\rightarrow j}$, which intuitively will include tasks of $P_i$ ``interfering'' with a $\mathit{Wait}(i, j)$ synchronization. In Figure~\ref{fig:lift-events}, task $B$ is interfering with the ``$E$ waits for $A$'' synchronization, therefore it is placed on a new stream $P_{1\rightarrow 0}$, and $E$ interferes with the ``$C$ waits for $D$'' event, therefore it is placed on a new stream $P_{0\rightarrow 1}$. We let original streams maintain their indices in the system. To map each new stream to a unique natural number index, we apply a Cantor pairing such that $P_{i\rightarrow j} = P_{m-1 + (i+j)(i+j+1)/2 +j}$, where $m$ is the number of original streams (number of PEs). Algorithmically, we iterate over tasks in time order and, for each task in their sync neighborhood, identify interfering tasks and update their stream. Note that we must introduce new synchronizations between the lifted task and the tasks before and after it in its original stream.
For example, in Figure~\ref{fig:lift-events}, $\mathit{Wait}(1,2)$ and $\mathit{Wait}(2,1)$ are introduced to enforce the order $A, B, C$. Overall, this approach is general and eliminates the effect demonstrated in Figure~\ref{fig:naive-events-limitation}, which  otherwise incurs slowdowns of up to $4\%$ in our Workload I experiments. All results of Workloads I and II make use of it to achieve the presented speedups. Adding new synchronization events in this setting incurs negligible runtime overhead in our experiments.

\subsection{Discussion}
\label{sec:experiments-discussion}
\paragraph{Summary of the comparison.} The methods in \cite{marchal2018parallel, bathie2021dynamic} are based on maximal topological cuts, which may limit their scalability. The state-of-the-art approach of \citet{fradet2025parallel}, while being an algorithmic method like ours, depends on the precomputation of a memory-optimal sequential schedule. The latter, although tractable for some instances \cite{fradet2023sequential,fradet2025graph}, is generally intractable and bound to turn into a bottleneck for  complex or very large, static or dynamic, inputs in contemporary applications. The experiments of Section~\ref{sec:experiments-scientific-workflows} show that our algorithm alone, without a precomputed sequential schedule, comes close to \mastag~in memory on a single PE, and schedules almost all instances on multiple PEs when the memory limit leaves some slack above the sequential peak of \lctes: $99.7\%$ to $100\%$ under M$+25\%$ to M$+75\%$. Already under M$+1\%$, which exceeds M by $0.4\%$ on median, it schedules $95\%$ to $96\%$ of them. Also, under the G limits, which come from our own sequential schedule, it schedules all instances, whereas \mastag~returns none on $5\%$ to $6.5\%$ of them, all cycles graphs, where \lctes~stops at its time limit or returns no schedule. A single run of our algorithm can also be executed online, at a loss in feasibility, see Table~\ref{tab:online}, whereas the orders of \mastag~need the whole graph. Our method may yield slightly worse memory than \mastag, and it fails under the tightest limit on about a third of the instances. In exchange, it schedules the largest graphs, where \lctes~does not return within a day, in less than a minute on a single PE. On the deep learning graphs both tools schedule, with \mastag~run without its refinement, the makespans of the two tools are within $1\%$ of each other on geometric mean and our portfolio is two orders of magnitude faster, as it does not need \lctes, see Section~\ref{sec:experiments-deep-learning}. The short running time of our algorithm makes it suitable for applications where scheduling time matters, e.g., rapid parameter tuning and many (iterative) compilations.

\paragraph{Limitations.} First, in Section~\ref{sec:experiments-scientific-workflows}, we select the portfolio on the same instances we report; the leave-one-workflow-out tests give an estimate of the loss on new workflows, about 4 points of worst-case feasibility and $0.4\%$ in makespan. Second, \mastag~exists only in \texttt{Python}, so we compare it with the \texttt{Python} implementation of our algorithm, on one machine. \texttt{C++} implementations of both would be faster, and the ratio between them might change. Third, \lctes~is limited to 600 seconds per call of its branch and bound and \mastag~to one day per graph, and each parallel scheduler of \mastag, like each method of our portfolio, to 600 seconds per call. With more time, \lctes~could find better schedules where it stops early, and the parallel schedulers of \mastag~could find shorter schedules on montage and cycles. Fourth, as in \cite{fradet2025parallel}, we do not charge \mastag~for the unconstrained list schedule its scheduler (iii) needs, which slightly favours \mastag~in running time. Fifth, the G limits come from our own sequential schedule, which favours our algorithm under these limits. Hence, we report feasibility mainly for the M limits. Finally, the deep learning results come from one framework and, for the training runs, one type of accelerator.

\section{Related Work}\label{sec:related}
\paragraph{List scheduling.} Our algorithm belongs to the family of greedy list scheduling algorithms. In list scheduling, tasks are sorted by a priority rule and each task is assigned to a processor, e.g., the one where it can start or finish the earliest, without revisiting earlier decisions. Graham's bound \cite{graham1969bounds} states that any such schedule on identical processors has makespan at most $2 - 1/m$ times the optimum. Many priority rules have been compared, based on costs, bottom and top levels, or critical paths \cite{adam1974comparison,shirazi1990analysis,ahmad1996analysis,kwok1999static}, and HEFT \cite{topcuoglu2002performance} extends list scheduling to heterogeneous processors, see also \cite{AlEbrahim2017TaskScheduling}. With processing set restrictions, i.e., when each task may only run on some processors, as in our deep learning setting, makespan bounds for list scheduling with typed tasks are given in \cite{liu1978performance,jaffe1980bounds}, see also the survey \cite{leung2008scheduling}. These algorithms do not consider memory. Under a memory limit, the makespan guarantee of list scheduling no longer holds, and a greedy schedule may not even exist; this is the setting we address. Our algorithm keeps the list scheduling structure, and adds a memory profile over the whole time horizon,  and, with \wspp, priorities that adapt to the schedule built so far.

\paragraph{Memory-aware scheduling.} For trees, makespan and memory are studied in \cite{herrmann2013model,marchal2013scheduling}, and list scheduling heuristics that respect a memory limit, by booking memory in advance, are given in \cite{eyraud2015parallel}. For general task graphs and a shared memory, fictitious edges are added in \cite{marchal2018parallel,marchal2019limiting} so that every parallel schedule respects a memory limit, and the maximum memory of any schedule on $p$ processors is computed in \cite{bathie2021dynamic}. Bounds on the parallel execution time under a memory limit are given in \cite{song2019toward}, and the worst-case memory over all $p$-processor executions of a fork-join program is computed in \cite{kaler2020cilkmem}.
Further memory-aware scheduling is studied in \cite{herrmann2014memory}, with two types of resources, each with its own memory, and in \cite{arras2015list}, where dedicated allocation and release tasks are included. In \cite{wang2022reservation}, memory-constrained reservation-based list scheduling is proposed, where memory is reserved for all successors up to depth $k$. In \cite{yao2022memory}, a quintic-time heuristic is proposed, including a memory-aware method for processor selection, when local memories and an unbounded shared memory are available. In \cite{li2023s}, a memory-aware depth-first search, where nodes with lower memory consumption are prioritized in ties, is proposed in a data materialization context. On heterogeneous processors, each with its own memory, workflows are partitioned and mapped in \cite{kulagina2024mapping}, and scheduled by memory-aware variants of HEFT with data eviction, checking memory at placement time, in \cite{kulagina2025memory}. In \cite{ding2023heuristic}, tasks are assigned to heterogeneous cores and data to memories of different speeds, with a slow memory that is unbounded in their experiments, and a tabu search improves a greedy list schedule; there, memory affects time but not feasibility.
In \cite{fradet2025parallel}, parallel schedules of scientific workflows are derived that respect the memory of an optimal sequential schedule. The sequential schedule is computed by graph transformations followed by branch and bound \cite{fradet2023sequential,fradet2025graph}. We compare to this approach in Section~\ref{sec:experiments-scientific-workflows}. 
In \cite{papp2025multiprocessor}, an integer linear program is used for memory-aware scheduling in a bulk synchronous parallel setting, where each processor has a bounded fast memory and a shared slow memory is unbounded. Most of the above works, like ours, bound or minimize the peak memory. In \cite{benoit2026minimizing}, the average memory consumption over time is minimized instead, which matters when several applications share the memory; optimal algorithms are given for some classes of task graphs under two memory models, and heuristics are built on them. Taking the average memory into account in our algorithm is left for future work.

\paragraph{Scheduling in deep learning.} In machine learning systems, scheduling mostly targets the overlap of computation and communication: tensor transfers are ordered or partitioned \cite{hashemi2019tictac,jayarajan2019priority}, communication partitioning or fusion is combined with scheduling \cite{chen2024centauri,cheng2025concerto}, and overlap is tuned together with memory-saving techniques at the level of training configurations \cite{zhu2025mist}. Scheduling individual operators under a memory bound, as we do, is not the focus of these works.

Further machine learning systems literature focuses on memory minimization, see Definition~\ref{def:graph-schedule-memory}, where all tasks execute on a single PE. Example approaches include brute force \cite{liberis2019neural}, exponential-time dynamic programming \cite{ahn2020ordering}, hierarchical operator organization  \cite{wang2022hierarchical}, integer linear programming \cite{zhong2023memory,steiner2023model,shu2023roam}, and learning \cite{gagrani2022neural}.

\section{Conclusion}\label{sec:conclusion}
We design a fast memory-aware list scheduling algorithm for task graphs on heterogeneous processors with a shared memory: it runs in polynomial time and returns either a schedule which provably respects the memory limit, or none. On the 339 of 341 scientific workflow task graphs where the sequential solver of \mastag~returns a schedule, our task sorting method \wspp~comes within $0.01\%$ of its memory while being $5.5$ times faster, on median. On 4 and 8 processors, without a precomputed sequential schedule, a small portfolio of variants schedules $63\%$ of the instances under a memory limit equal to the sequential peak of \mastag, and $95\%$ to $96\%$ under one that exceeds it by less than $0.4\%$ on median. The portfolio is, on median, $7$ to $12$ times faster than \mastag, and $2.9$ to $4.1$ times faster without the optional refinement of \mastag, with makespans within $0.2\%$ to $8.1\%$ of those of \mastag~over all limits we test. In MindSpore, the algorithm gives training speedups up to $1.10\times$ over a manually fine-tuned schedule with at most $3.62\%$ more memory, and memory savings up to $18\%$. On seven deep learning graphs with up to 115,528 tasks, our \texttt{Python} implementation, with \wspp, computes a sequential schedule in at most 33 seconds. With the time limit of its experiments, the sequential solver of \mastag~takes 13 minutes to 2.2 hours on five of them and returns none within a day on the two largest.

In the future, we plan to evaluate the algorithm on more general heterogeneous platforms, and to improve its robustness under tight memory limits, both for the static algorithm, e.g., with a backtracking component when scheduling time is less critical, and for its online variant. We may also take the average memory over time into account, as in \cite{benoit2026minimizing}. Another direction is to combine our algorithm with \mastag, e.g., to seed the parallel schedulers of \mastag~with our sequential order instead of the one of \lctes.

\bibliographystyle{plainnat}
\bibliography{references}

\end{document}